\documentclass[notitlepage,nofootinbib,preprintnumbers,amssymb,superscriptaddress]{revtex4-1}

\usepackage{amsfonts,amssymb,amsmath,graphicx,color,bm,multirow}
\usepackage[utf8]{inputenc}
\definecolor{ultramarine}{rgb}{0.07, 0.04, 0.56}
\definecolor{cadmiumgreen}{rgb}{0.0, 0.42, 0.24}
\definecolor{indigo(dye)}{rgb}{0.0, 0.25, 0.42}
\usepackage[dvipdfmx]{hyperref}
\hypersetup{
colorlinks=true,
citecolor=ultramarine,
linkcolor=cadmiumgreen,
urlcolor=indigo(dye),
}

\newcommand{\meff}{m_{\mathrm{eff}}}
\newcommand{\Veff}{V_{\mathrm{eff}}}
				
\newcommand{\rhoPT}{\rho_{\mathrm{PT}}}	
			
\newcommand{\RNS}{R_{\mathrm{NS}}}
\newcommand{\MNS}{M_{\mathrm{NS}}}

\newcommand{\mui}{\mu_{\mathrm{i}}}
\newcommand{\nui}{\nu_{\mathrm{i}}}
\newcommand{\ri}{r_{\mathrm{i}}}
\newcommand{\pii}{\tilde{p}_{\mathrm{i}}}
\newcommand{\phii}{\phi_{\mathrm{i}}}
\newcommand{\psii}{\psi_{\mathrm{i}}}

\newcommand{\pf}{\tilde{p}_{\mathrm{f}}}
\newcommand{\phif}{\phi_{\mathrm{f}}}
\newcommand{\psif}{\psi_{\mathrm{f}}}

\begin{document}


\title{Spontaneous scalarization with an extremely massive field \\and heavy neutron stars}

\author{Soichiro Morisaki}
\affiliation{Research Center for the Early Universe (RESCEU), Graduate School of Science, The University of Tokyo, Tokyo 113-0033, Japan}
\affiliation{Department of Physics, Graduate School of Science, The University of Tokyo, Tokyo 113-0033, Japan}
\author{Teruaki Suyama}
\affiliation{Research Center for the Early Universe (RESCEU), Graduate School of Science, The University of Tokyo, Tokyo 113-0033, Japan}

\begin{abstract}
We investigate the internal structure and the mass-radius relation of neutron stars in a recently proposed scalar-tensor theory dubbed asymmetron in which a massive scalar field undergoes spontaneous scalarization inside neutron stars.
We focus on the case where the Compton wavelength is shorter than $10~\mathrm{km}$, which has not been investigated in the literature.
By solving the modified Einstein equations, either purely numerically or by partially using a semianalytic method, we find that not only the weakening of gravity by spontaneous scalarization but also the scalar force affect the internal structure significantly in the massive case.
We also find that the maximum mass of neutron stars is larger for certain parameter sets than that in general relativity and reaches $2 M_{\odot}$ even if the effect of strange hadrons is taken into account.
There is even a range of parameters where the maximum mass of neutron stars largely exceeds the threshold that violates the causality bound in general relativity. 
\end{abstract}

\maketitle

\section{Introduction}\label{sec:intro}
General relativity (GR) has been tested by various experiments \cite{Will:2014kxa}. 
Especially, the precise tests in the Solar system put stringent constraints on the deviation from GR.
In fact, GR is a nonrenormalizable theory in quantum field theory and cannot be applied beyond the Planck scale.
In this sense, it should be regarded as an effective field theory of an as-yet-unknown quantum gravity theory.
However, even at the classical level of the effective field theory, the correct theory may not be GR.
Actually, the existence of dark matter and dark energy may suggest the modification of GR in the extreme regime not probed by the terrestrial or solar-system experiments, and 
various theories which pass the solar-system experiments and accommodate the dark sector have been proposed \cite{Joyce:2014kja}. 
In order to test these possibilities, it is important to test gravity beyond the currently accessible scale.

Recently, gravitational waves emitted from a binary black hole coalescence were detected for the first time \cite{Abbott:2016blz} and two signals as well as a less significant candidate followed it \cite{TheLIGOScientific:2016pea,Abbott:2017vtc}. 
Needless to say, gravitational waves will become a new tool to test gravity in the near future.
In fact, the detected signals put constraints on the gravity theory in the highly dynamical and strong-field regime for the first time \cite{TheLIGOScientific:2016src,TheLIGOScientific:2016pea,Yunes:2016jcc}. 
Although these new constraints are weaker than the existing ones for most of the theories, more precise tests will be possible as the detector sensitivity will be improved.

One of the main targets of the gravitational-wave observations, which is also relevant to our present study, is neutron stars. 
A neutron star is a highly dense star, whose density reaches the nuclear density. 
There is not enough evidence to show that GR is correct in such a highly dense region.
On the other hand, the structure of neutron stars in various theories of gravity has been investigated by many authors (See \cite{Berti:2015itd} and references therein. \cite{Babichev:2016jom,Sakstein:2016oel,Minamitsuji:2016hkk,Maselli:2016gxk,Aoki:2016eov,Cisterna:2015yla,Cisterna:2016vdx} are recent work on this subject not included in Ref. \cite{Berti:2015itd}.). 
Since the structure varies from that in GR, it is possible to distinguish these theories by observations of neutron stars in principle.
However, things are not so trivial because of the degeneracy between the uncertainties in the equation of state for nuclear matter and gravitational physics.
One way to break this degeneracy is to use equation-of-state-independent relations \cite{Postnikov:2010yn,Hinderer:2009ca,Yagi:2013bca}.
The I-Love-Q relation \cite{Yagi:2013bca} is one such relation, which holds between the moment of inertia, I, the Love number and the quadrupole moment, Q. 
The Love number can be measured by gravitational-wave observations in the near future \cite{Flanagan:2007ix}.
Therefore, tests of gravity with neutron stars will become feasible in the coming era.

Another topic related to neutron stars is the existence of massive neutron stars, whose masses are about 2$M_{\odot}$ \cite{Demorest:2010bx,Antoniadis:2013pzd}. Especially, PSR J1614-2230 \cite{Demorest:2010bx} shows a strong Shapiro delay signature, and the estimated mass of the pulsar turned out to be $M=(1.97 \pm 0.04)M_{\odot}$. This measurement does not assume any models of emission mechanism of the stars and the result is robust. On the other hand, although the state of nuclear matter inside the highly dense core of the neutron stars is unknown, it is natural that strange hadrons appear there since the chemical potential of the neutrons is large enough. 
For example, it is pointed out that hyperons appear when the baryon number density, $n$, surpasses the threshold density of $(2-3)n_0$ \cite{Nishizaki:2002ih}, where $n_0=0.17~\mathrm{nucleons/fm^3}$ is the saturation density.
Such strange hadrons soften the equation of state and significantly reduces the maximum mass of the neutron stars that can be supported against gravity. 
A 2$M_{\odot}$ neutron star is difficult to realize with these strange hadrons (see Fig. 3 of \cite{Demorest:2010bx}). 
In order to solve this potential inconsistency, many ways to improve the equations of state for nuclear matter have been suggested, such as taking into account three-baryon interactions \cite{Nishizaki:2002ih,Gandolfi:2015jma,RikovskaStone:2006ta} or a quark matter core \cite{Bonanno:2011ch,Masuda:2012ed}.
On the other hand, it is also possible that modification of GR solves this inconsistency \cite{Astashenok:2014pua}.

One simple extension of GR is scalar-tensor theory \cite{Fujii:2003pa}.
In this theory, a new scalar degree of freedom is added to GR.
This additional degree mediates a new force called the scalar force and it affects the motion of particles.
For example, it modifies the trajectory of light around the Sun from that in GR. 
Currently, the measurement of the time delay of light with the Cassini spacecraft \cite{Bertotti:2003rm} puts stringent constraints on the coupling constant between the scalar field and matter, and the scalar force is negligible in the Solar System.

Damour and Esposito-Farese proposed an interesting scalar-tensor theory which passes the constraints and modifies the structure of neutron stars significantly \cite{Damour:1993hw}. 
In this theory the value of the scalar field in the Solar System is so small that the coupling constant, which is zero when the scalar field is vanishing, is negligibly small. Therefore, this theory safely passes Solar System experiments.
On the other hand, inside neutron stars, the scalar field gets a much larger value than in the Solar System and the deviation from GR is significant.
This phenomena is called spontaneous scalarization \cite{Damour:1996ke}.
Especially, the effective gravitational constant decreases and gravity is weakened in the spontaneous-scalarization phase. 
Therefore, this theory may allow more massive neutron stars than in GR and solve the aforementioned problem of the observed existence of massive neutron stars.

The model proposed by Damour and Esposito-Farese has two problems. 
The first is that this model is already severely constrained by binary pulsar observations \cite{Antoniadis:2013pzd,Damour:1996ke}.
Especially, from the observation of PSR J0348-0432 \cite{Antoniadis:2013pzd}, it was found that the effect of spontaneous scalarization must be negligibly small even inside the massive neutron star of the binary, whose mass is $(2.01\pm0.04)M_{\odot}$.
The second is that spontaneous scalarization can occur during the inflation and matter dominated eras.
Therefore, the scalar field gets a large value in the present universe for a broad range of its initial value, which means fine-tuning for the initial value is necessary for this theory to pass solar-system experiments \cite{Sampson:2014qqa,Damour:1992kf,Damour:1993id}.

These problems do not exist if the scalar field is massive \cite{Chen:2015zmx}.
First, if the Compton wavelength of the scalar field, $\lambda_\phi$, is much smaller than the periapse of the orbit of PSR J0348-0432, which is the order of $10^{10}~\mathrm{m}$ \cite{Antoniadis:2013pzd}, the scalar field does not affect the orbital motion of the binary and the test with this binary is safely passed \cite{Ramazanoglu:2016kul}.
In addition, even if spontaneous scalarization occurs in the early universe, the scalar field begins damped oscillation when the Hubble parameter becomes equal to the mass and converges to zero without fine tuning \cite{Chen:2015zmx}.
In other words, GR is a cosmological attractor in the late time universe.
Therefore, models of spontaneous scalarization with a massive scalar field have recently attracted attention and been investigated.
In \cite{Ramazanoglu:2016kul}, the structure of neutron stars in the case where $\lambda_\phi \gtrsim 100~\mathrm{km}$ was investigated. 
On the other hand, the structure of neutron stars in the case of the shorter Compton wavelength has not been investigated.
This region is interesting since the oscillating component of the scalar field may account for the overall amount of dark matter. This scenario is dubbed asymmetron scenario \cite{Chen:2015zmx}.

In this paper, we investigate the structure of neutron stars in the model of spontaneous scalarization with a massive scalar field introduced in \cite{Chen:2015zmx}, focusing on the case where $\lambda_\phi<10~\mathrm{km}$. We also investigate the maximum mass of neutron stars and identify the parameter space where massive neutron stars exceeding $2M_{\odot}$ are allowed with strange hadrons. The rest of the paper is organized as follows. In Sec. \ref{sec:basic}, we explain our model and the basic equations. In Sec. \ref{sec:method}, we explain the numerical method we employ to obtain the structure of neutron stars. In Sec. \ref{sec:result}, we show the results of our analysis, that is, the internal structure, the mass-radius relation, and the maximum mass of neutron stars in our model. The last section is devoted to the conclusion.


\section{Basic equations}\label{sec:basic}
We briefly review the model of the massive scalar field proposed in \cite{Chen:2015zmx} and
the basic equations we use in our analysis.
\subsection{Model}
As in \cite{Chen:2015zmx}, we introduce a real massive scalar field $\phi$ in the gravitational sector, whose Compton wavelength $\lambda_\phi \equiv 1/m_\phi$, where $m_\phi$ is the mass of $\phi$, is smaller than $10~\mathrm{km}$.
We assume that this scalar field couples with ordinary matter fields $\Psi_m$ universally through the physical metric $\tilde{g}_{\mu \nu} \equiv A^2(\phi) g_{\mu \nu}$, which is called the Jordan metric (compared to this metric, $g_{\mu \nu}$ is called the Einstein metric.). This coupling guarantees that the weak equivalence principle holds true. We consider the action given by
\begin{align}
S&=S_g[g_{\mu \nu},\phi] +S_m[\tilde{g}_{\mu \nu},\Psi_m] \\
&=\int d^4 x \sqrt{-g} \left( \frac{R}{16 \pi G} - \frac{1}{2} g^{\mu \nu} \partial_\mu \phi \partial_\nu \phi - \frac{m^2_\phi}{2} \phi^2 \right)+\int d^4 x \sqrt{-\tilde{g}} \mathcal{L}_m (\tilde{g}_{\mu \nu},\Psi_m),
\end{align}
where $G$ is the gravitational constant measured in laboratory experiments and $\mathcal{L}_m$ is the Lagrangian of all the other matter fields. 
The equations of motion derived from this action are the following:
\begin{align}
&G_{\mu \nu} = 8 \pi G \left[ - \left(\frac{1}{2} g^{\alpha \beta} \partial_\alpha \phi \partial_\beta \phi + \frac{1}{2} m^2_\phi \phi^2 \right) g_{\mu \nu} + \partial_\mu \phi \partial_\nu \phi + A^2(\phi) \tilde{T}_{\mu \nu} \right], \label{modEinstein} \\
&\Box_g \phi - m^2_\phi \phi +\alpha A^4 \tilde{T} = 0, \label{scalarEOM}
\end{align}
where $\alpha(\phi)$ is defined by $\alpha(\phi) \equiv d\mathrm{ln}A(\phi)/d \phi$.
$\tilde{T}_{\mu \nu}$ and $\tilde{T}$ are the stress-energy tensor and its trace defined by the Jordan metric respectively, that is, 
\begin{equation}
\tilde{T}_{\mu \nu} \equiv -\frac{2}{\sqrt{-\tilde{g}}} \frac{\delta S_m}{\delta \tilde{g}^{\mu \nu}},~~~\tilde{T} \equiv \tilde{g}^{\mu \nu} \tilde{T}_{\mu \nu}.
\end{equation}
According to Eq. (\ref{modEinstein}), the effective gravitational constant, $A^2(\phi)G$, changes when the value of $A^2(\phi)$ deviates from $1$.
The equation of motion for the scalar field $\phi$ can be rewritten as
\begin{equation}
\Box_g \phi - \frac{d \Veff}{d \phi} = 0,~~~\Veff \equiv \frac{1}{2}m^2_\phi \phi^2-\frac{1}{4} \tilde{T} A^4(\phi).
\end{equation}
This shows that dynamics of $\phi$ is described by an effective potential $\Veff (\phi)$, which depends on the density of surrounding matter.
As is clear from the expression of $\Veff$, if the second derivative of $A$ at the origin is negative, the spontaneous scalarization happens when $-\tilde{T}$ exceeds a critical value \cite{Chen:2015zmx}.

In this study, the scalar force plays an important role.
The strength of the scalar force is characterized by $\alpha(\phi)$.
As an example, we consider two point particles interacting with each other only through $\tilde{g}_{\mu \nu}$ in vacuum. We assume that the background value of $\phi$ is $0$, which is the stable point of the potential $V(\phi)$.
In the Newtonian limit, the equation of motion for the particles is
\begin{equation}
\frac{d^2 \vec{x}}{d t^2} = -\vec{\nabla}(\Phi_\mathrm{g}+\Phi_\mathrm{s}), \label{newton_law}
\end{equation}
where 
\begin{equation}
\Phi_\mathrm{g}=-G\frac{M_{\mathrm{par}}}{r},~~~\Phi_{\mathrm{s}}=-\frac{\alpha^2(0)}{4 \pi}\frac{M_{\mathrm{par}}}{r}\mathrm{e}^{-m_{\phi}r},
\end{equation}
and $M_{\mathrm{par}}$ is the mass of the other particle.
The first term in the right-hand side of Eq. (\ref{newton_law}) corresponds to gravitational force and the second corresponds to the scalar force.
It can be easily seen that the absolute value of $\alpha(0)$ controls the strength of the scalar force.

\begin{figure}
	\begin{center}
		\includegraphics[width = 12cm]{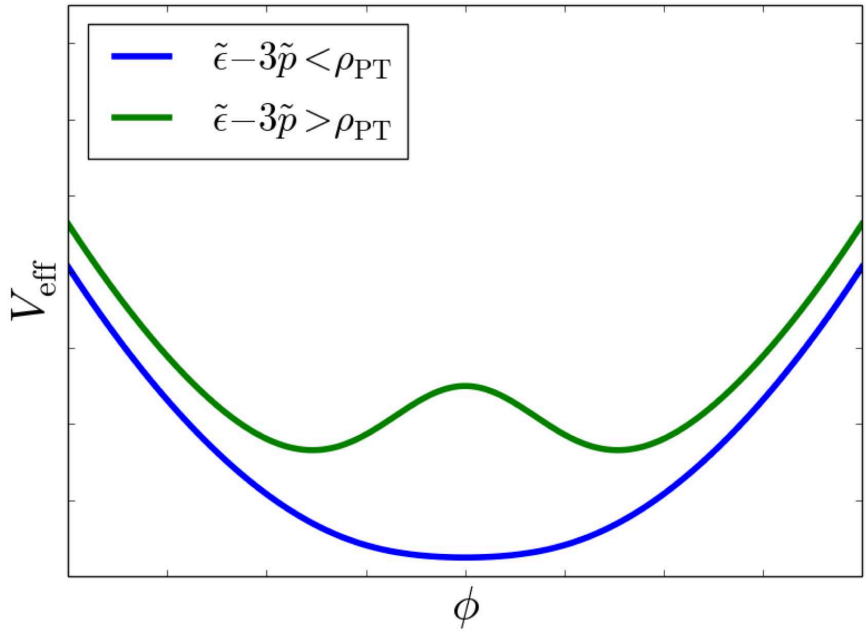}
		\caption{The shape of the effective potential $\Veff(\phi)$. For $\tilde{\epsilon}-3 \tilde{p}<\rhoPT$, which is realized in the Solar System, it has a stable point at $\phi=0$. For $\tilde{\epsilon}-3 \tilde{p} > \rhoPT$, which is realized inside neutron stars in the case of our study, $\phi=0$ becomes unstable and nontrivial stable points appear.}
	\label{effective}
	\end{center}
\end{figure}

Following \cite{Chen:2015zmx}, we use
\begin{equation}
A^2(\phi)=1-\eta+\eta \mathrm{exp}\left(-\frac{\phi^2}{2 M^2}\right),~~0<\eta \leq 1
\end{equation}
throughout our study. In our study we assume that the matter is perfect fluid and $-\tilde{T}=\tilde{\epsilon}-3 \tilde{p}$, where $\tilde{\epsilon}$ is energy density and $\tilde{p}$ is pressure in the Jordan frame. Therefore the effective potential is
\begin{align}
\Veff (\phi) &=\frac{1}{2} m^2_\phi \phi^2 + \frac{1}{4}(\tilde{\epsilon}-3 \tilde{p}) \left(1-\eta+\eta \mathrm{exp}\left(-\frac{\phi^2}{2 M^2}\right) \right)^2\\
& =\frac{1}{4}(\tilde{\epsilon}-3 \tilde{p})+\frac{1}{2} \left(m^2_\phi - \frac{\eta (\tilde{\epsilon}-3 \tilde{p})}{2 M^2} \right)\phi^2 + \mathcal{O}(\phi^4).
\end{align}
It can be easily seen that the point $\phi=0$ becomes unstable and spontaneous symmetry breaking occurs if $\tilde{\epsilon}-3 \tilde{p} > \rhoPT$, where the critical density $\rhoPT$ is defined by
\begin{equation}
\rhoPT \equiv \frac{2 m^2_\phi M^2}{\eta}.
\end{equation}
The shape of the effective potential $\Veff (\phi)$ is given in Fig. \ref{effective}. The value of $\rhoPT$ is assumed to be somewhat smaller than mass density of a nucleon, $\rhoPT \leq \rho_{\mathrm{nuc}} \sim 10^{15}~\mathrm{g/cm^3}$. In this case, spontaneous symmetry breaking occurs only inside neutron stars, as we explicitly demonstrate later. 
In the Solar System, where $\tilde{\epsilon}-3 \tilde{p} \simeq \tilde{\rho} \simeq 1~ \mathrm{g/cm^3} \ll \rhoPT$, $\phi$ vanishes everywhere.
Accordingly, $\alpha$ also vanishes and the general relativity is classically recovered.
Therefore, our model safely passes the solar-system tests. On the other hand, inside a neutron star, where $\tilde{\epsilon}-3 \tilde{p} >\rhoPT$, $\Veff$ has stable points at
\begin{equation}
\phi=\pm \bar{\phi},~~\bar{\phi} \equiv M\sqrt{2\mathrm{ln}\left[\frac{2 \eta}{1-\eta}\left(\sqrt{1+\frac{4\eta\rhoPT}{(1-\eta)^2 (\tilde{\epsilon}-3\tilde{p})}}-1\right)^{-1}\right]}. \label{phitop}
\end{equation}
In this region, $\phi$ stays near the nontrivial stable point, which is $\mathcal{O}(M)$, and $A^2(\phi)<1$. Therefore, the effective gravitational constant, $A^2(\phi) G$, becomes smaller than that in the low density region and gravitational force is weakened inside neutron stars.

\subsection{Modified Einstein equations for a static and spherical star}
Throughout our study we neglect spins of neutron stars and consider a static and spherically symmetric configuration, that is, 
\begin{equation}
ds^2 = g_{\mu \nu} dx^{\mu} dx^{\nu} = -\mathrm{e}^{\nu(r)} dt^2 + \frac{dr^2}{1-\frac{2 \mu(r)}{r}} + r^2 d \Omega^2,~\phi=\phi(r),~\tilde{p}=\tilde{p}(r),~\tilde{\epsilon}=\tilde{\epsilon}(r).
\end{equation}
The field equations Eq. (\ref{modEinstein}) and Eq. (\ref{scalarEOM}) lead to
\begin{align}
\mu'&=2 \pi G(r(r-2\mu)\psi^2+r^2 m^2_\phi \phi^2)+4 \pi G A^4(\phi) r^2 \tilde{\epsilon}, \label{mueq}\\
\nu'&=4 \pi G r \psi^2 + \frac{1}{r(r-2\mu)}(8 \pi G r^3 A^4(\phi) \tilde{p}-4 \pi G r^3 m^2_\phi \phi^2 + 2\mu),\label{nueq}\\
\tilde{p}'&=-\frac{\tilde{\epsilon}+\tilde{p}}{2}(\nu'+2 \alpha \psi), \label{peq}\\
(r - 2 \mu)\psi' &= -2\left(1-\frac{\mu}{r}\right)\psi + m^2_\phi(4 \pi G r^2 \phi^2 \psi+r \phi)+r A^4(\phi) (4 \pi G r(\tilde{\epsilon}-\tilde{p})\psi+\alpha(\tilde{\epsilon}-3\tilde{p})), \label{psieq}\\
\phi'&=\psi, \label{phieq}
\end{align}
where $'$ denotes differentiation with respect to the radial coordinate $r$. We have introduced a new variable $\psi \equiv \phi'$ in order to reduce the equations to the first order differential equations. Equation (\ref{peq}) is the hydrostatic equilibrium condition for nuclear matter. In addition to the standard gravitational force, the scalar force described by the second term also contributes to the equilibrium condition. As we will see later, this scalar force significantly changes the internal structure of neutron stars. The surface of the star, $r=\RNS$, is defined as the surface on which $\tilde{p}$ becomes $0$, and $\tilde{p}$ is set to be $0$ outside the star. 

Next, we discuss boundary conditions. Nonsingular solutions satisfy the following conditions at the origin:
\begin{equation}
\mu(0)=0,~\nu(0)=0,~\tilde{p}(0)=\tilde{p}_c,~\psi(0)=0. \label{initial}
\end{equation} 
Outside the star, $\phi$ has an exponentially growing solution and an exponentially decaying solution. Since the former one is physically unacceptable, we impose another boundary condition at infinity as
\begin{equation}
\lim_{r \to \infty} \phi(r) =0. \label{boundary}
\end{equation}
This condition is equivalent to choosing a suitable value of $\phi$ at the origin.
For $r \gtrsim R_{NS}+\lambda_\phi$, $g_{\mu \nu}$ becomes almost identical to the physical metric $\tilde{g}_{\mu \nu}$.
Since $\lambda_\phi$ we are interested in is much shorter than the typical orbital distance of binary pulsars, the mass of the neutron star estimated by the binary pulsars measurement, $M_{NS}$, is equal to $\mu(\infty)/G$.

\begin{figure}
  \begin{center}
    \begin{tabular}{c}

      \begin{minipage}{0.5\hsize}
        \begin{center}
          \includegraphics[clip, width=8cm]{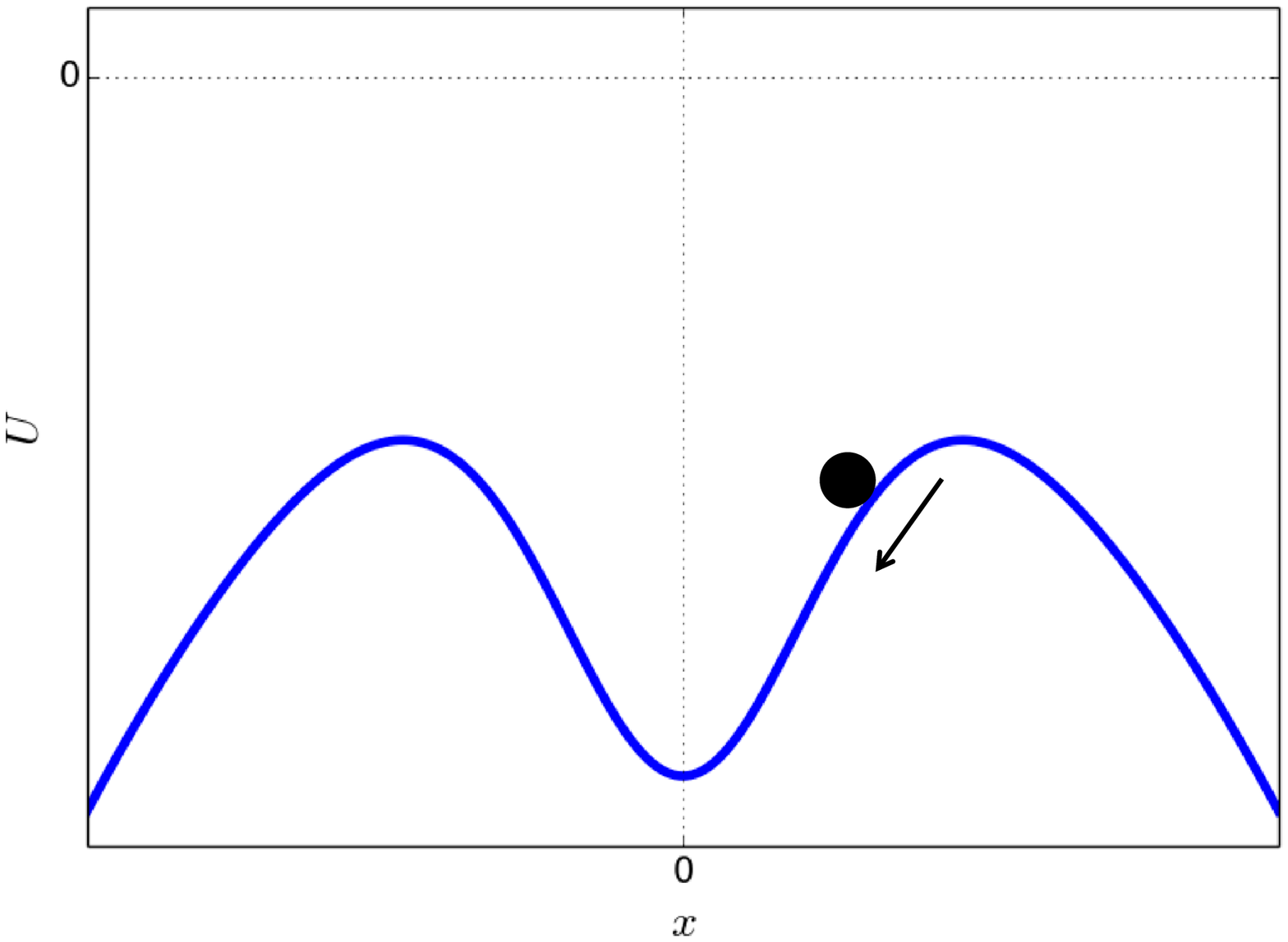}
           \end{center}
      \end{minipage}

      \begin{minipage}{0.5\hsize}
        \begin{center}
          \includegraphics[clip, width=8cm]{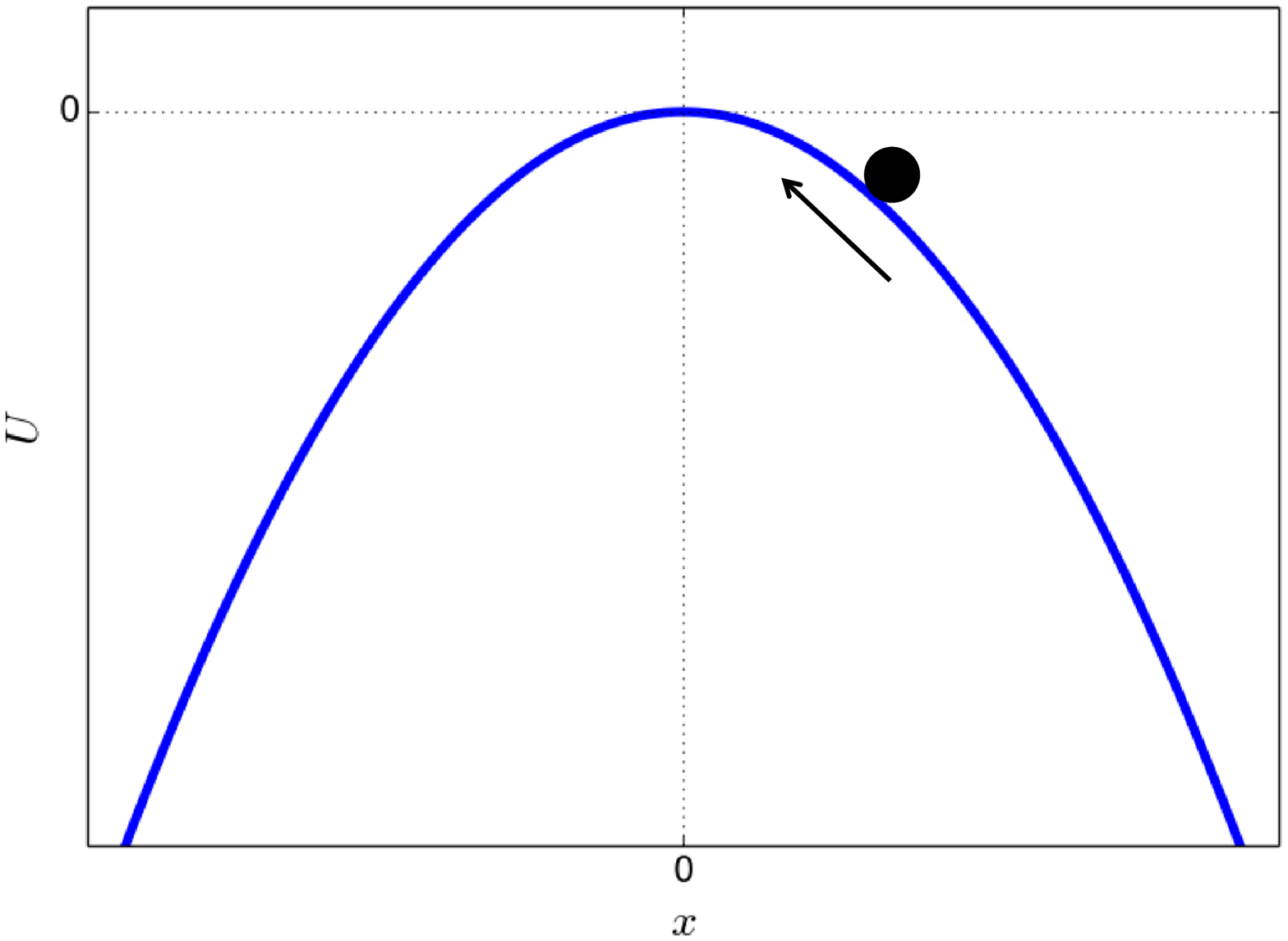}
            \end{center}
      \end{minipage} \\

    \end{tabular}
    \caption{The shape of $U$ and the motion of the particle corresponding to $\phi$ are shown for different densities. The left one shows the plot for $\tilde{\epsilon}-3 \tilde{p} > \rhoPT$ and the right one for $\tilde{\epsilon}-3 \tilde{p} < \rhoPT$. The black dot represents the particle and the arrow represents the direction of the motion.}
    \label{dynamical}
  \end{center}
\end{figure}

Finally, we explain roughly how $\phi(r)$ changes in $r$.
As explained in II. C of \cite{Chen:2015zmx}, ignoring the curvature of the spacetime and performing the change of variables as
\begin{equation}
r \rightarrow \tau,~\phi \rightarrow x,~\Veff \rightarrow -U,
\end{equation}
we can reduce Eq. (\ref{psieq}) and Eq. (\ref{phieq}) to
\begin{equation}
\frac{d^2 x}{d \tau^2} + \frac{2}{\tau} \frac{d x}{d \tau} =-\frac{ d U}{d x}.
\end{equation}
It is the same as the equation for a motion of a  particle under the potential $U$ with time dependent friction, and it helps us to understand the profile of $\phi(r)$ \cite{Khoury:2003rn}.
In the core of the star, where $\tilde{\epsilon}-3 \tilde{p}>\rhoPT$, the particle, $\phi$, rolls down from $\bar{\phi}$ (the left plot of Fig. \ref{dynamical}).
The potential $U$ around $x=\bar{\phi}$ can be approximated by
\begin{equation}
U=-\frac{1}{2} \meff^2 (x-\bar{\phi})^2
\end{equation}
with $\meff(\tilde{p}) \equiv \sqrt{V''_{\mathrm{eff}}(\bar{\phi}(\tilde{p}))}$.
For $\tilde{\epsilon}-3 \tilde{p} \gg \rhoPT$, $\meff$ is 
\begin{equation}
\meff \simeq m_\phi \sqrt{2 \mathrm{ln}\left(\frac{(1-\eta)(\tilde{\epsilon}-3 \tilde{p})}{\rhoPT}\right)} \label{meff_est}
\end{equation}
and of the order of $m_\phi$.
\footnote{For $\eta=1$, Eq. (\ref{meff_est}) can not be applied. However $\meff$ can be written as $\meff=m_\phi \sqrt{2\mathrm{ln}[(\tilde{\epsilon}-3 \tilde{p})/\rhoPT]}$ and is of the order of $m_\phi$ even in this case.}
Therefore, the deviation from $\bar{\phi}$ increases exponentially with the scale of $\mathcal{O}(\lambda_\phi)$.
After the density decreases and $\tilde{\epsilon} - 3\tilde{p}$ becomes lower than $\rhoPT$, the potential U is convex around $x=0$.
According to the boundary condition equation (\ref{boundary}), $\phi$ climbs this convex potential and reaches $\phi=0$ at infinity (the right plot of Fig \ref{dynamical}). 

\subsection{Equations of state}
In addition to the equations from Eq. (\ref{mueq}) to Eq. (\ref{phieq}), we need an equation of state. We adopt an equation of state defined in the Jordan frame,
\begin{equation}
\tilde{\epsilon}=\tilde{\epsilon}(\tilde{p}), \label{EOS}
\end{equation}
since it is the one that is directly provided by the nuclear physics experiments and theory. In our study, we consider two types of equations of state, that for hadronic matter and for strange quark matter. The latter is the matter composed of deconfined up, down and strange quarks. Although whether stars consisting of strange quark matter exist or not has not been clarified, their formation path and properties have been discussed in the literature \cite{Itoh:1970uw,Witten:1984rs,Haensel:1986qb,Alcock:1986hz,Olinto:1986je}.

\subsubsection{Hadronic matter}
To model equations of state for hadronic matter, we use piecewise polytropic equations, which are constructed by a few polytropic equations \cite{Read:2008iy}. A polytropic equation is in the form of 
\begin{equation}
\tilde{p}(\tilde{\rho}) = \bar{p}\left(\frac{\tilde{\rho}}{\bar{\rho}}\right)^{\Gamma},
\end{equation}
where $\tilde{\rho}$ is the rest-mass density of the matter and $\Gamma$ is the adiabatic index. Integrating the first law of thermodynamics
\begin{equation}
d\left(\frac{\tilde{\epsilon}}{\tilde{\rho}}\right)=-\tilde{p} d\left( \frac{1}{\tilde{\rho}} \right), \label{1stlaw}
\end{equation}
we obtain the relation in the form of Eq. (\ref{EOS}) as
\begin{equation}
\tilde{\epsilon}(\tilde{p})=(1+a)\bar{\rho}\left(\frac{\tilde{p}}{\bar{p}}\right)^{\frac{1}{\Gamma}}+\frac{\tilde{p}}{\Gamma-1},
\end{equation}
where $a$ is an integration constant. Piecewise polytropic equations are constructed by connecting polytropic equations at each density contained in a set of dividing densities $\rho_1<\rho_2< \dots <\rho_N$. For each interval the equation of state is 
\begin{equation}
\tilde{p}(\tilde{\rho})=p_i \left(\frac{\tilde{\rho}}{\rho_i}\right)^{\Gamma_i},~~\tilde{\epsilon}(\tilde{p})=(1+a_i)\rho_i\left(\frac{\tilde{p}}{p_i}\right)^{\frac{1}{\Gamma_i}}+\frac{\tilde{p}}{\Gamma_i-1},~~\rho_{i-1} \leq \tilde{\rho} < \rho_i
\end{equation} 
and
\begin{equation}
\tilde{p}(\tilde{\rho})=p_N \left(\frac{\tilde{\rho}}{\rho_N}\right)^{\Gamma_{N+1}},~\tilde{\epsilon}(\tilde{p})=(1+a_{N+1})\rho_N \left(\frac{\tilde{p}}{p_N} \right)^{\frac{1}{\Gamma_{N+1}}}+\frac{p}{\Gamma_{N+1}-1},~~\tilde{\rho} \geq \rho_N
\end{equation}
with $\rho_0=0$. In the limit $\tilde{p} \to 0$ the matter becomes nonrelativistic, that is, $\tilde{\epsilon} \to \tilde{\rho}$. It leads to
\begin{equation}
a_1 =0.
\end{equation}
The continuity condition for $\tilde{p}$ and $\tilde{\epsilon}$ at $\tilde{\rho}=\rho_i$ leads to
\begin{align}
p_{i+1}&=p_i \left( \frac{\rho_{i+1}}{\rho_i} \right)^{\Gamma_{i+1}}~~(i=1,2,\dots,N-1),\\
a_{i+1}&=a_i+\left(\frac{1}{\Gamma_i-1}-\frac{1}{\Gamma_{i+1}-1}\right)\frac{p_i}{\rho_i}~~(i=1,2,\dots,N).
\end{align}
Therefore, the equation of state is characterized by $p_1$, $\Gamma_i~(i=1,2, \dots ,N+1)$ and $\rho_i~(i=1,2,\dots,N)$.
In our study we use two $N=6$ piecewise polytropes, which approximate AP4 (which contains only $\mathrm{npe\mu}$ as nuclear matter) \cite{Akmal:1998cf} and GS1 (which contains kaons, which are strange hadrons, in addition to $\mathrm{npe\mu}$) \cite{Glendenning:1997ak}. 
The GS1 is one of the equations of state taking into account the effect of strange hadrons and the maximum mass for it in GR is $1.37 M_{\odot}$, which does not reach the measured mass of the neutron star PSR J1614-2230.
It is one of the goals of this work to clarify whether the maximum mass for the GS1 reaches this measured value in our model.
Following \cite{Read:2008iy}, we use the piecewise polytrope approximating the SLy equation of state at low density and three polytropes at high density for both cases.
The parameters for both of the equations of state at low density and at high density are  separately listed in Tables \ref{polytrope1} and \ref{polytrope2}.

\begin{table}[h]
	\caption{We use the four polytropes defined by the parameters listed below at much lower density than nuclear density.   }
	\begin{center}
	\begin{tabular}{cc}
	\begin{minipage}[t]{.45\textwidth}
		\begin{center}
			\begin{tabular}{|c|c|} \hline
				$p_1$ & $3.08879\times 10^{24}~ \mathrm{dyn/cm^2}$ \\ 
				$\Gamma_1$ & $1.58425$ \\ 
				$\rho_1$ & $2.44034\times 10^7~\mathrm{g/cm^3}$ \\
				$\Gamma_2$ & $1.28733$ \\ \hline
			\end{tabular}
		\end{center}
	\end{minipage}
	
	\hfill
	
	\begin{minipage}[t]{.45\textwidth}
		\begin{center}
			\begin{tabular}{|c|c|} \hline
				$\rho_2$ & $3.78358\times 10^{11}~\mathrm{g/cm^3}$ \\ 
				$\Gamma_3$ & $0.62223$ \\ 
				$\rho_3$ & $2.62780 \times 10^{12}~\mathrm{g/cm^3}$ \\
				$\Gamma_4$ & $1.35692$ \\ \hline
			\end{tabular}
		\end{center}
	\end{minipage}
	\end{tabular}
	\label{polytrope1}
	\end{center}				
\end{table}

\begin{table}[h]
\begin{center}
\caption{We use the three polytropes defined by the parameters listed below around nuclear density. The dividing densities $\rho_5$ and $\rho_6$ are common for both of the equations of state. }
\begin{tabular}{|c|c|c|c|c|c|c|} \hline
$~$& $\rho_4$ & $\Gamma_5$ & $\rho_5$ & $\Gamma_6$ & $\rho_6$ & $\Gamma_7$ \\ \hline
AP4 & $1.51200 \times 10^{14}~\mathrm{g/cm^3}$ & 2.830 & \multirow{2}{*}{$5.01187 \times 10^{14}~\mathrm{g/cm^3}$} & 3.445 & \multirow{2}{*}{ $10^{15}~\mathrm{g/cm^3}$} & 3.348 \\ \cline{2-3} \cline{5-5} \cline{7-7}
GS1 & $4.91314 \times 10^{13}~\mathrm{g/cm^3}$ & 2.350 &  & 1.267 & & 2.421 \\ \hline
\end{tabular}
\label{polytrope2}
\end{center}
\end{table}

\subsubsection{Strange quark matter}
The simplest model of strange quark matter, which we use, is the MIT bag model \cite{Chodos:1974je}. Its equation of state is
\begin{equation}
\tilde{\epsilon}=3 \tilde{p} + 4 B,~~~B \simeq 56~\mathrm{MeV/fm^3},
\end{equation}
where we have chosen $B$ to be $3/8$ of the saturation density $\epsilon_\mathrm{s}\simeq150~\mathrm{MeV/fm^3}$, following \cite{Lattimer:2012nd}.


\section{Method to solve the equations}\label{sec:method}
In this section, we explain the numerical methods we use in this study. We adopt different methods for $10~\mathrm{km} \geq \lambda_\phi \gtrsim 1~\mathrm{km}$ (mildly massive case) and $\lambda_\phi \ll 1~\mathrm{km}$ (very massive case).

\subsection{$10~\mathrm{km} \geq \lambda_\phi \gtrsim 1~\mathrm{km}$ case (mildly massive case)}
We numerically integrate the equations (\ref{mueq}) - (\ref{phieq}) outwards from the center using the Runge-Kutta-Fehlberg method, imposing the initial conditions  Eq. (\ref{initial}) and
\begin{equation}
\phi(0)=\phi_c.
\end{equation}
Because some terms in the equations are apparently singular at the origin $r=0$ and can not be handled with by numerical computations, we use the Taylor expansion of the solution around $r=0$ to start the computations at $r \neq 0$. 

In order to satisfy the boundary condition Eq. (\ref{boundary}), the value of the scalar field at the origin $\phi_c$ must be tuned to a certain value. We use shooting method to find such physical solutions. We look for the appropriate value for $\phi_c$ in the interval between $\phi=0$ and a sufficiently large value by means of bisection search. The detail of the bisection search is as follows.
\begin{enumerate}
\item If we observe $\phi(r)$ grows exponentially to positive direction, we take smaller value of $\phi_c$ for the next trial.
\item If we observe $\phi(r)$ becomes negative, we take larger value of $\phi_c$ for the next trial\footnote{We assume that $\phi(r)$ tracks the stable point of $\Veff$ and is positive everywhere for correct $\phi_c$}.
\item After repeating this procedure many times, we obtain $\phi(r)$ which shows convergent behavior at sufficiently large r. We stop the numerical integration if the integration reaches the point $r=r_c$ in the vacuum at which the following conditions are satisfied.
\begin{equation}
r \gg \lambda_\phi,~4 \pi G r^2 \phi^2 \ll 1,~4 \pi G r^2 \phi^2 \ll m_\phi \mu. \label{WKB_condition}
\end{equation}
The third condition guarantees that the contribution from the energy of the scalar field at $r>r_c$ to $\mu(\infty)$ is small enough to be ignored.
Around that point, the following WKB solution can be applied  (The derivation of the WKB solution is summarized in Appendix \ref{sec:appendixA}).
\begin{equation}
\phi(r)=\frac{C_+}{r\left(1-\frac{2 \mu}{r} \right)^{\frac{1}{4}}} \mathrm{exp}\left(\int^{r}_{r_c} \frac{m_{\phi}}{\sqrt{1-\frac{2 \mu}{r}}} dr \right) + \frac{C_-}{r\left(1-\frac{2 \mu}{r} \right)^{\frac{1}{4}}}\mathrm{exp}\left(-\int^{r}_{r_c} \frac{m_{\phi}}{\sqrt{1-\frac{2 \mu}{r}}} dr \right). \label{WKB1}
\end{equation}
We connect the numerical solution with the WKB solution and calculate $C_+$ and $C_-$. If $|C_+/C_-|<1$, the deviation from the correct configuration of $\phi(r)$ arising from the incorrectness of the value of $\phi_c$ is small all over the region, $r<r_c$. Therefore, we stop the bisection search and adopt the resultant configuration as a good approximate solution. We approximate the mass of the neutron star as $\mu(r_c)/G$. On the other hand, if $|C_+/C_-|>1$, we take larger value of $\phi_c$ for the next trial if $C_+<0$ and vice versa. 
\end{enumerate}

\subsection{$\lambda_\phi \ll 1~\mathrm{km}$ case (very massive case)}\label{subsec:method2}
If $\lambda_\phi \ll 1~\mathrm{km}$, it is very difficult to solve the equations numerically all over the space because the required accuracy for the numerical integration is extremely high. For example, if the numerical error in $\phi$ arises at $r=r_0$, it  increases by a factor of
\begin{equation}
\mathrm{exp}\left[C \frac{r_c-r_0}{\lambda_\phi} \right]
\end{equation}
up to $r_c$, where $C$ is a $\mathcal{O}(1)$ coefficient. $C$ represents the difference between $m^{-1}_{\mathrm{eff}}$ and $\lambda_\phi$, and $C \sim 3$ for $\eta=1,~\rhoPT=10^{8}~\mathrm{MeV^4},~\tilde{\epsilon}-3 \tilde{p} = 10^{10}~\mathrm{MeV^4}$. Since $r_c \sim \RNS \sim 10~\mathrm{km}$, the numerical error arising near the origin increases by a factor of 
\begin{equation}
\mathrm{exp}\left[3 \left(\frac{10~\mathrm{km}}{\lambda_\phi}\right) \right] \sim 10^{13\left(\frac{1~\mathrm{km}}{\lambda_\phi}\right)}.
\end{equation}
Therefore, the necessary precision is too high to achieve, and we have to develop an alternative method to obtain an approximate solution in the limit, $\lambda_\phi \ll 1~\mathrm{km}$, which we will explain below.

\begin{figure}
	\begin{center}
		\includegraphics[width = 12cm]{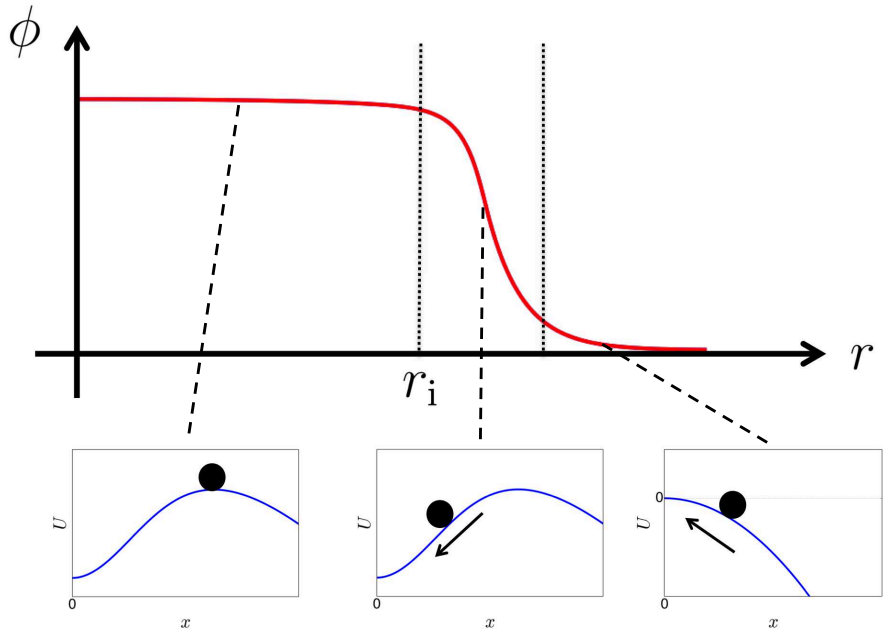}
		\caption{The schematic profile of $\phi(r)$. We can separate it into three regions. The detail is explained in the text.}
	\label{schematic}
	\end{center}
\end{figure}

We can model the profile of $\phi(r)$ as shown in Fig. \ref{schematic}. As pointed out in II. C of \cite{Chen:2015zmx}, $\phi$ stays extremely close to $\bar{\phi}$ up to a certain radius in this regime. Then, $\phi$ leaves $\bar{\phi}$ and transitions toward $0$ somewhere inside the star with the scale of $\meff^{-1}=\mathcal{O}(\lambda_\phi)$. In the transition region, the scalar force becomes significant and surpasses gravitational force, that is,
\begin{equation}
2 \alpha \psi \gg \nu'. \label{surpass}
\end{equation}
This inequality can be easily confirmed as follows. 
According to Eq. (\ref{nueq}), noticing $\mu$ is larger than the other terms in the right side because it is the integrated value of the energy density from the center, we can estimate $\nu'$ by
\begin{equation}
\nu' \simeq \frac{2 \mu}{r^2}.
\end{equation}  
Since $r>2 \mu$ inside neutron stars, we have
\begin{equation}
\nu' \lesssim \frac{\mathcal{O}(1)}{r}.
\end{equation}
On the other hand, the scalar force can be estimated by
\begin{equation}
2 \alpha \psi = 2 \frac{d \mathrm{ln} A(\phi)}{d r} \simeq \frac{\mathcal{O}(1)}{\lambda_\phi}
\end{equation}
since $\mathrm{ln}A$ increases by $\mathcal{O}(1)$ with the scale of $\mathcal{O}(\lambda_\phi)$.
Since we now consider extremely short Compton wavelength, we can assume $\lambda_\phi \ll r$ in this region. Therefore, Eq. (\ref{surpass}) holds. Notice that $r$ is typically $\RNS$ in the present case and the effect of the scalar force becomes significant when $\lambda_\phi \ll \RNS$.
Since the scalar force compresses the star significantly, the pressure decreases significantly in this region.
For later convenience, we define $r=\ri$ as the radial coordinate of the point where the scalar force becomes comparable to the scalar force, that is,
\begin{equation}
2 \alpha(\phii) \psii = \nu'(\ri), \label{rs_def}
\end{equation}
where the subscript i means the value at $r=\ri$.
After the roll-down phase, $\phi$ climbs up the mountain of $U$ and reaches $\phi=0$ at infinity.

Next, we explain the detail of each region. 
In the region, $r<\ri$, $\phi$ remains extremely close to $\bar{\phi}$. Therefore, we can use the approximation,
\begin{equation}
\phi(r) \simeq \bar{\phi}(\tilde{p}(r)). \label{phi=phibar}
\end{equation}
In addition, we have 
\begin{equation}
\psi(r) \ll m_\phi \bar{\phi}(\tilde{p}(r)) \label{psi<<mphi} 
\end{equation}
since $\bar{\phi}$ changes with much longer scale than $\lambda_\phi$. The justification of Eq. (\ref{phi=phibar}) and Eq. (\ref{psi<<mphi}) is explained in Appendix \ref{sec:appendixB}.
Therefore, we can approximate  Eq. (\ref{mueq}), Eq. (\ref{nueq}) and Eq. (\ref{peq}) as
\begin{align}
\mu'&=2 \pi G r^2 m^2_\phi \bar{\phi}(\tilde{p})^2+4 \pi G A^4(\bar{\phi}(\tilde{p})) r^2 \tilde{\epsilon}, \label{mueq2}\\
\nu'&=\frac{1}{r(r-2\mu)}(8 \pi G r^3 A^4(\bar{\phi}(\tilde{p})) \tilde{p}-4 \pi G r^3 m^2_\phi \bar{\phi}(\tilde{p})^2 + 2\mu), \label{nueq2}\\
\tilde{p}'&=-\frac{(\tilde{\epsilon}+\tilde{p})\nu'}{2\left(1+(\tilde{\epsilon}+\tilde{p})\alpha(\bar{\phi})\frac{d \bar{\phi}}{d \tilde{p}}\right)}. \label{peq2}
\end{align}
By integrating these equations numerically, we can obtain an approximate solution for this region.
The important point is that we do not have to solve Eq. (\ref{psieq}) and Eq. (\ref{phieq}).
Therefore, we have no numerical difficulties in this integration.

In the region where $\phi$ rolls down on the length scale $\lambda_\phi$, we can approximate Eq. (\ref{peq}) as
\begin{equation}
\frac{d \tilde{p}}{d r} =-(\tilde{\epsilon}+\tilde{p}) \frac{d \mathrm{ln} A}{d r}\label{peq3}
\end{equation}
since the scalar force becomes dominant.
In addition, since
\begin{equation}
\ri \gg \lambda_\phi, \label{r<<l}
\end{equation}
the region where the scalar force is dominant is an extremely thin shell and we can treat $r,~\mu$ as constants, that is,
\begin{equation}
r=\ri,~\mu=\mui. \label{constants}
\end{equation}
With Eq. (\ref{r<<l}) and Eq. (\ref{constants}), we can approximate Eq. (\ref{psieq}) and Eq. (\ref{phieq}) as
\begin{equation}
\frac{d^2 \phi}{d r^2}=\frac{1}{1 - \frac{2 \mui}{\ri}} \left(m^2_\phi \phi + \alpha A^4(\phi)(\tilde{\epsilon}- 3 \tilde{p})\right). \label{psieq3}
\end{equation}
Equation (\ref{peq3}) and Eq. (\ref{psieq3}) can be analytically integrated once. 
We first consider the hadronic equations of state and explain the case of the strange quark matter after that. 
In this case, the integration from $\ri$ to $r$ leads to
\begin{align}
\frac{\tilde{\epsilon}+\tilde{p}}{\tilde{\rho}} &= \frac{\tilde{\epsilon}_\mathrm{i}+\pii}{\tilde{\rho}_\mathrm{i}}\frac{A(\bar{\phi}(\tilde{p}_i))}{A(\phi)}, \label{p_integrate}\\
\frac{1}{2}\psi^2 &= \frac{1}{1-\frac{2 \mui}{\ri}} \left[\frac{1}{2}m^2_\phi (\phi^2-\bar{\phi}^2(\pii))+\pii A^4(\bar{\phi}(\pii))-\tilde{p} A^4(\phi)\right]. \label{phi_integrate}
\end{align}
The derivations of Eq. (\ref{psieq3}), Eq. (\ref{p_integrate}) and Eq. (\ref{phi_integrate}) are explained in Appendix \ref{sec:appendixC} in detail.

Since $\tilde{p}$ decreases significantly in this region, there are two cases to consider.
If 
\begin{equation}
\frac{\tilde{\epsilon}_\mathrm{i}+\pii}{\tilde{\rho}_\mathrm{i}} A(\bar{\phi}(\pii)) \leq 1 \label{case1}
\end{equation}
is satisfied, $\tilde{p}$ becomes $0$ while $\phi$ is rolling down. In this case, Eq. (\ref{phi_integrate}) leads to the following condition satisfied by the physical quantities at $r=\ri$ and on the surface:
\begin{equation}
\frac{1}{2}\psif^2 = \frac{1}{1-\frac{2 \mui}{\ri}} \left[\frac{1}{2}m^2_\phi (\phif^2-\bar{\phi}^2(\pii))+\pii A^4(\bar{\phi}(\pii)) \right], \label{condition1}
\end{equation}
where the subscript f means the value on the surface in this case. 
On the other hand, if
\begin{equation}
\frac{\tilde{\epsilon}_\mathrm{i}+\pii}{\tilde{\rho}_\mathrm{i}} A(\bar{\phi}(\pii)) > 1 \label{case2}
\end{equation}
is satisfied, $\tilde{p}>0$ even after $\phi$ has rolled down. 
In this case, Eq. (\ref{p_integrate}) and Eq. (\ref{phi_integrate}) lead to
\begin{align}
\frac{\tilde{\epsilon}_{\mathrm{f}}+\pf}{\tilde{\rho}_\mathrm{f}} &= \frac{\tilde{\epsilon}_{\mathrm{i}}+\pii}{\tilde{\rho}_{\mathrm{i}}}A(\bar{\phi}(\pii)), \label{p_condition2}\\
\frac{1}{2}\psif^2 &= \frac{1}{1-\frac{2 \mui}{\ri}} \left[\frac{1}{2}m^2_\phi (\phif^2-\bar{\phi}^2(\pii))+\pii A^4(\bar{\phi}(\pii))-\pf \right], \label{phi_condition2}
\end{align}
where the subscript f means the value at the position where $\phi/M\ll1$ is satisfied and the scalar force becomes negligible.

For both cases, $\phi$ climbs the mountain afterwards, and $\phi(r)$ in this phase can be approximated by the following WKB solution:
\begin{equation}
\phi(r)=C \mathrm{exp}\left( - \int^r \frac{\meff}{\sqrt{1-\frac{2\mu}{r}}} dr\right). \label{phi_vac}
\end{equation}
$\meff =m_\phi$ for the case of Eq. (\ref{case1}) and $\meff \simeq m_\phi \sqrt{1-(\tilde{\epsilon}-3 \tilde{p})/\rhoPT}$ for the case of Eq. (\ref{case2}). 
Substituting Eq. (\ref{phi_vac}) into Eq. (\ref{condition1}) and Eq. (\ref{phi_condition2}), we obtain the equations to determine $\pii$. 
As a result, we find that $\pii$ can be obtained as a zero point of $f(\pii)$, which is defined as follows:
\begin{equation}
f(\pii)= \begin{cases}
\pii A^4 (\bar{\phi}(\pii)) - \frac{m^2_\phi}{2} \bar{\phi}^2(\pii) & \frac{\tilde{\epsilon}_\mathrm{i}+\pii}{\tilde{\rho}_\mathrm{i}}A(\bar{\phi}(\pii)) \leq1 \\
\pii A^4 (\bar{\phi}(\pii)) - \frac{m^2_\phi}{2} \bar{\phi}^2(\pii)-\pf   & \frac{\tilde{\epsilon}_\mathrm{i}+\pii}{\tilde{\rho}_\mathrm{i}}A(\bar{\phi}(\pii)) >1,\end{cases}
\end{equation}
where $\pf$ is determined by Eq. (\ref{p_condition2}).
For the MIT bag model, $f(\pii)$ is defined as follows:
\begin{equation}
f(\pii)= \begin{cases}
\pii A^4 (\bar{\phi}(\pii)) - \frac{m^2_\phi}{2} \bar{\phi}^2(\pii) & (\pii + B)A^4(\bar{\phi}(\pii)) \leq B \\
B(1-A^4(\bar{\phi}(\pii)))-\frac{m^2_\phi}{2}\bar{\phi}^2(\pii)  &  (\pii + B)A^4(\bar{\phi}(\pii)) >B \end{cases}
\end{equation}

Obtaining $\pii$, we can obtain an approximate solution as follows.
First, we integrate Eq. (\ref{mueq2}), Eq. (\ref{nueq2}) and Eq. (\ref{peq2}) from $r=0$ to $r=\ri$, where $\tilde{p}=\pii$.
Then if $\pii$ satisfies Eq. (\ref{case1}), we stop the integration since $\tilde{p}$ becomes $0$ suddenly at that point.
$\RNS$ and $\MNS$ can be obtained as $\ri$ and $\mui/G$ respectively.
On the other hand, if $\pii$ satisfies Eq. (\ref{case2}), we have to continue the numerical integration from $r=\ri,~\mu(\ri)=\mui,~\nu(\ri)=\nui,~\tilde{p}(\ri)=\pf$. 
Since we can neglect the contribution of $\phi$, we just have to integrate the Einstein equations of GR up to the point where $\tilde{p}=0$.
$\RNS$ and $\MNS$ can be obtained as $r$ and $\mu/G$ at that point respectively.

\section{Result}\label{sec:result}
In this section, we explain the results of our analysis. We show the internal structure, mass-radius relationship and the maximum mass of neutron stars.

\subsection{The internal structure of neutron stars}
\begin{figure}
  \begin{center}
    \begin{tabular}{c}

      \begin{minipage}{0.5\hsize}
        \begin{center}
          \includegraphics[clip, width=8cm]{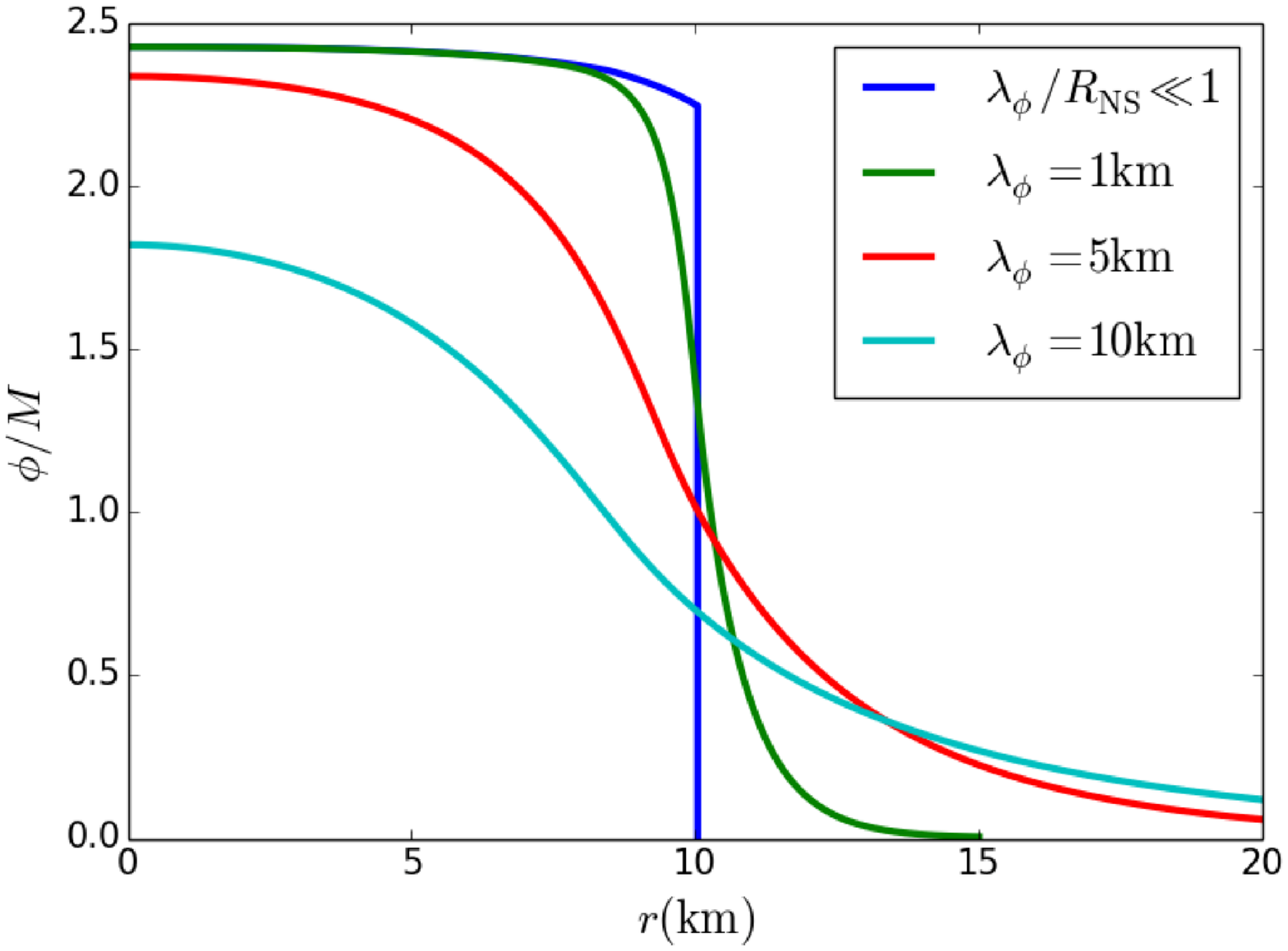}
           \end{center}
      \end{minipage}

      \begin{minipage}{0.5\hsize}
        \begin{center}
          \includegraphics[clip, width=8cm]{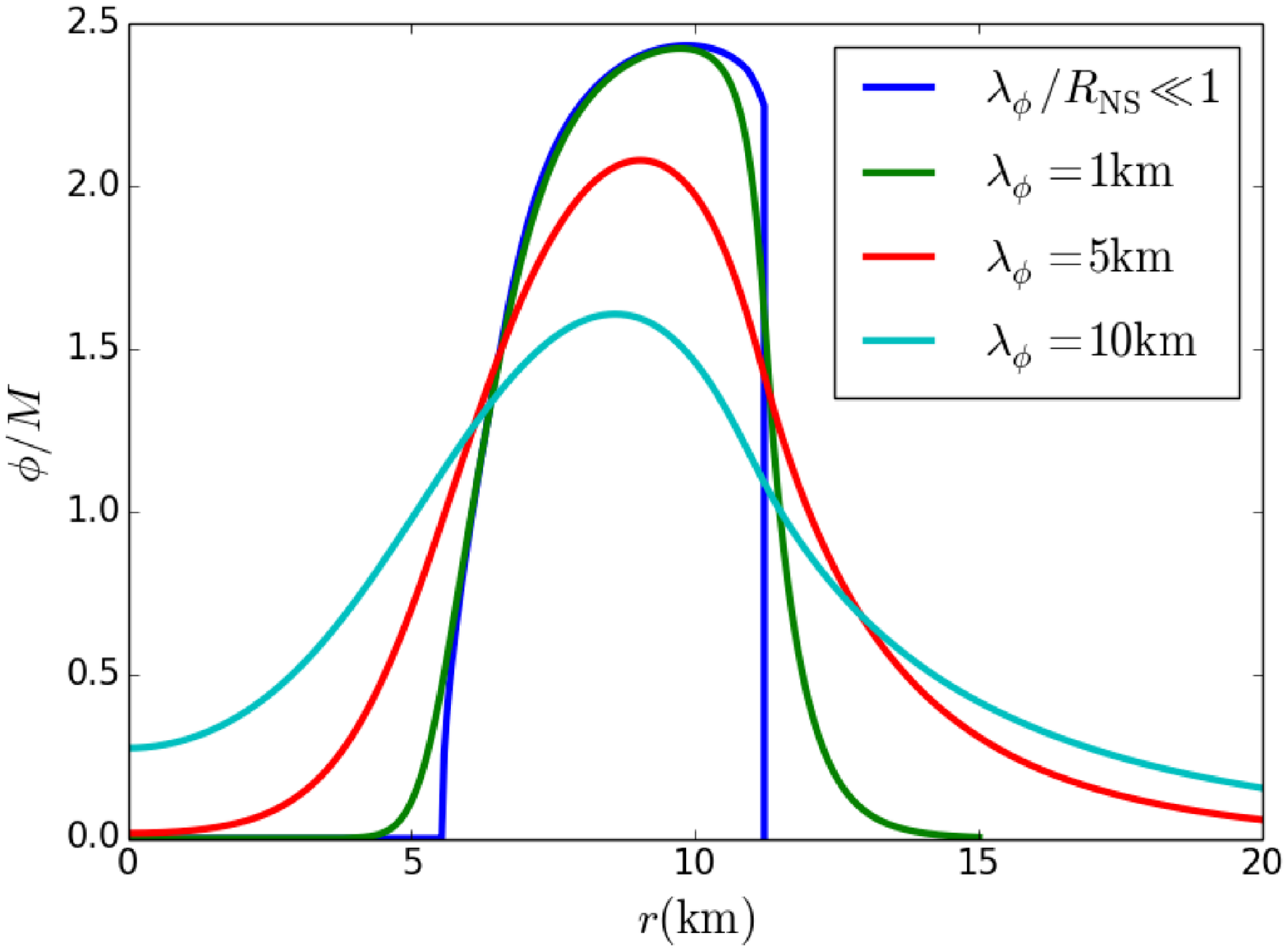}
            \end{center}
      \end{minipage} \\

    \end{tabular}
    \caption{The profiles of $\phi$ for various values of $\lambda_\phi$ are shown for each value of $\tilde{\rho}(r=0)$. The left one: $\tilde{\rho}(r=0) = 3 \times 10^{9} ~\mathrm{MeV^4}=7.0 \times 10^{14}~\mathrm{g/cm^3}$, the right one: $\tilde{\rho}(r=0)=7 \times 10^{9} ~\mathrm{MeV^4}=1.6 \times 10^{15}~\mathrm{g/cm^3}$. They have AP4 for the equation of state and $\eta=0.2,~\rhoPT=10^{8}~\mathrm{MeV^4}=2.3\times10^{13}~\mathrm{g/cm^3}$ for the values of the parameters in our model.}
    \label{scalar_profile}
  \end{center}
\end{figure}

The profiles of $\phi(r)$ for four different wavelengths and two different central densities are shown in Fig. $\ref{scalar_profile}$.
For each case, we observe that as $\lambda_\phi$ decreases the profile approaches what we obtain by means of the semianalytical method which we discuss in Sec. \ref{subsec:method2}.
This demonstrates the validity of our semianalytical method. We also observe that $\phi$ grows in the core  for the higher central density while $\phi$ is monotonic decreasing function of $r$ for the lower central density. It is because $-\tilde{T}=\tilde{\epsilon}-3 \tilde{p}$ is not a monotonically increasing function of $\tilde{\rho}$. For the hadronic equations of state and $\tilde{\rho}>10^{15}\mathrm{g/cm^3}$, $-\tilde{T}$ can be written as  
\begin{align}
-\tilde{T}&= \tilde{\epsilon}-3 \tilde{p} \nonumber \\
&=(1+a_{7}) \rho_6 \left(\frac{\tilde{p}}{p_6}\right)^{\frac{1}{\Gamma_{7}}}+\frac{4-3 \Gamma_7}{\Gamma_7-1}\tilde{p}.
\end{align}
Since typical hadronic equations have $\Gamma_7>4/3$, the second term becomes dominant and $-\tilde{T}$ turns to decrease as $\tilde{p}$, or $\tilde{\rho}$, increases.
Such a behavior is specific to the hadronic equations of state since $\tilde{\epsilon}-3 \tilde{p}=4 B = \mathrm{const.}$ and $-\tilde{T}$ is always positive for the MIT Bag model.
For the central density of the right plot of Fig. \ref{scalar_profile}, $-\tilde{T}<\rhoPT$ at $r=0$. For such high central densities and short Compton wavelengths, $\phi$ sits extremely close to $\phi=0$ in the core and starts to grow near the spherical surface, $r=r_{\mathrm{PT}}$, on which $-\tilde{T}=\rhoPT$ (See the green line of the right plot of Fig. \ref{scalar_profile}). 
Strictly speaking, the approximation, $\phi=\bar{\phi}$, is not valid at $r=r_{\mathrm{PT}}$, and this seems to imply that the semianalytic method explained in the previous section is no longer valid.
However, this is not true, as we explain below.
Since the behavior of the effective mass, $\meff$, near $r=r_{\mathrm{PT}}$ can be written as
\begin{equation}
\meff \simeq m_\phi \sqrt{\frac{2}{\rhoPT} \left.\frac{d}{dr}(\tilde{\epsilon}-3 \tilde{p})\right|_{r=r_{\mathrm{PT}}} (r-r_{\mathrm{PT}})},
\end{equation}
$\meff$ becomes large close to $r=r_{\mathrm{PT}}$ for large $m_\phi$.
On the other hand, $\phi$ must return back to $\bar{\phi}$ before $\meff$ becomes much larger than $1/\RNS$ since otherwise $\phi$ rolls down quickly and the boundary condition is not satisfied. 
Therefore, the region where the approximation, $\phi=\bar{\phi}$, is invalid is so thin that its effect on the internal structure is negligibly small, which means our semianalytical method is still applicable in this case.

\begin{figure}
  \begin{center}
    \begin{tabular}{c}

      \begin{minipage}{1\hsize}
        \begin{center}
          \includegraphics[clip, width=11.5cm]{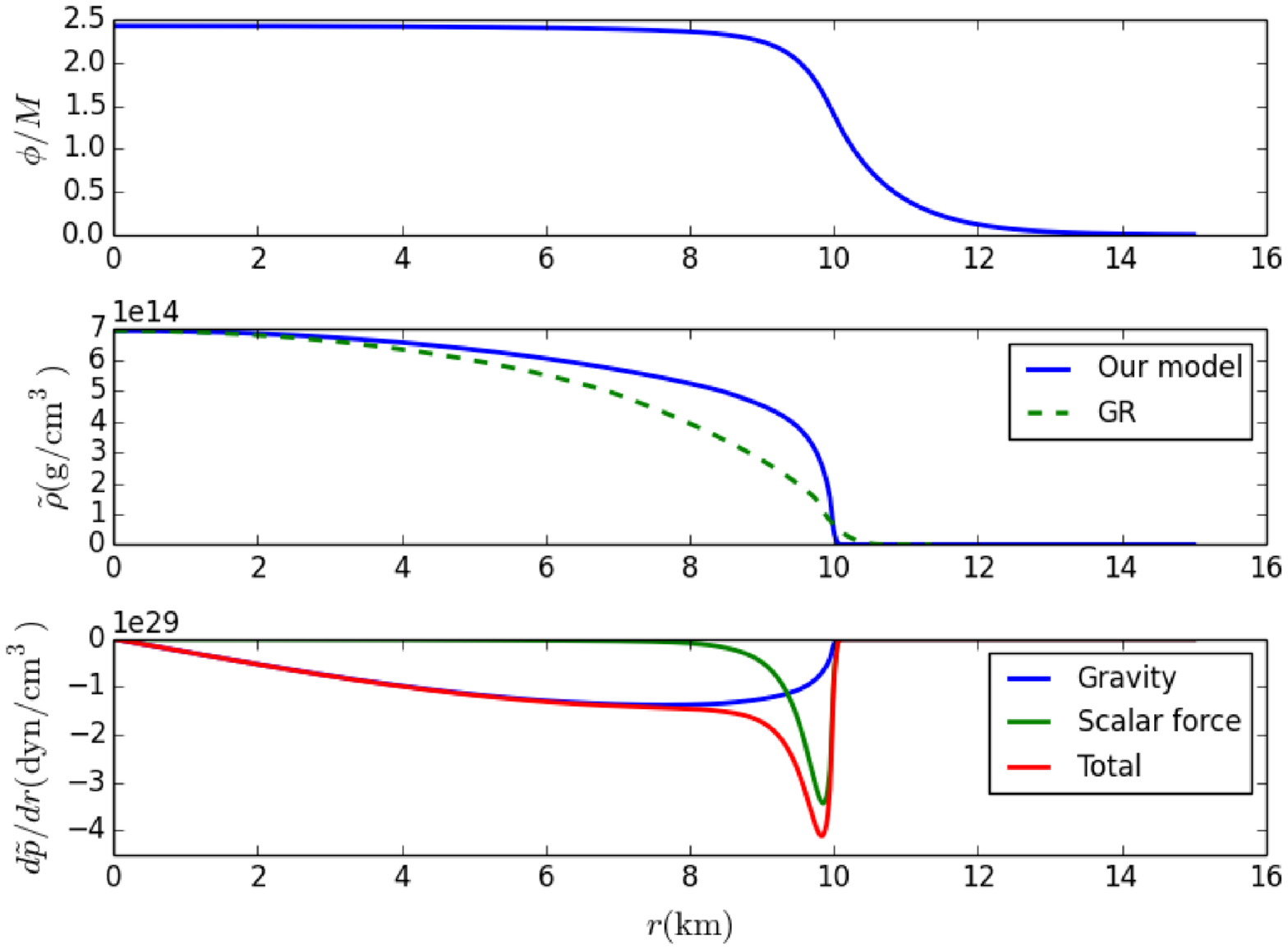}
           \end{center}
      \end{minipage}\\

      \begin{minipage}{1\hsize}
        \begin{center}
          \includegraphics[clip, width=11.5cm]{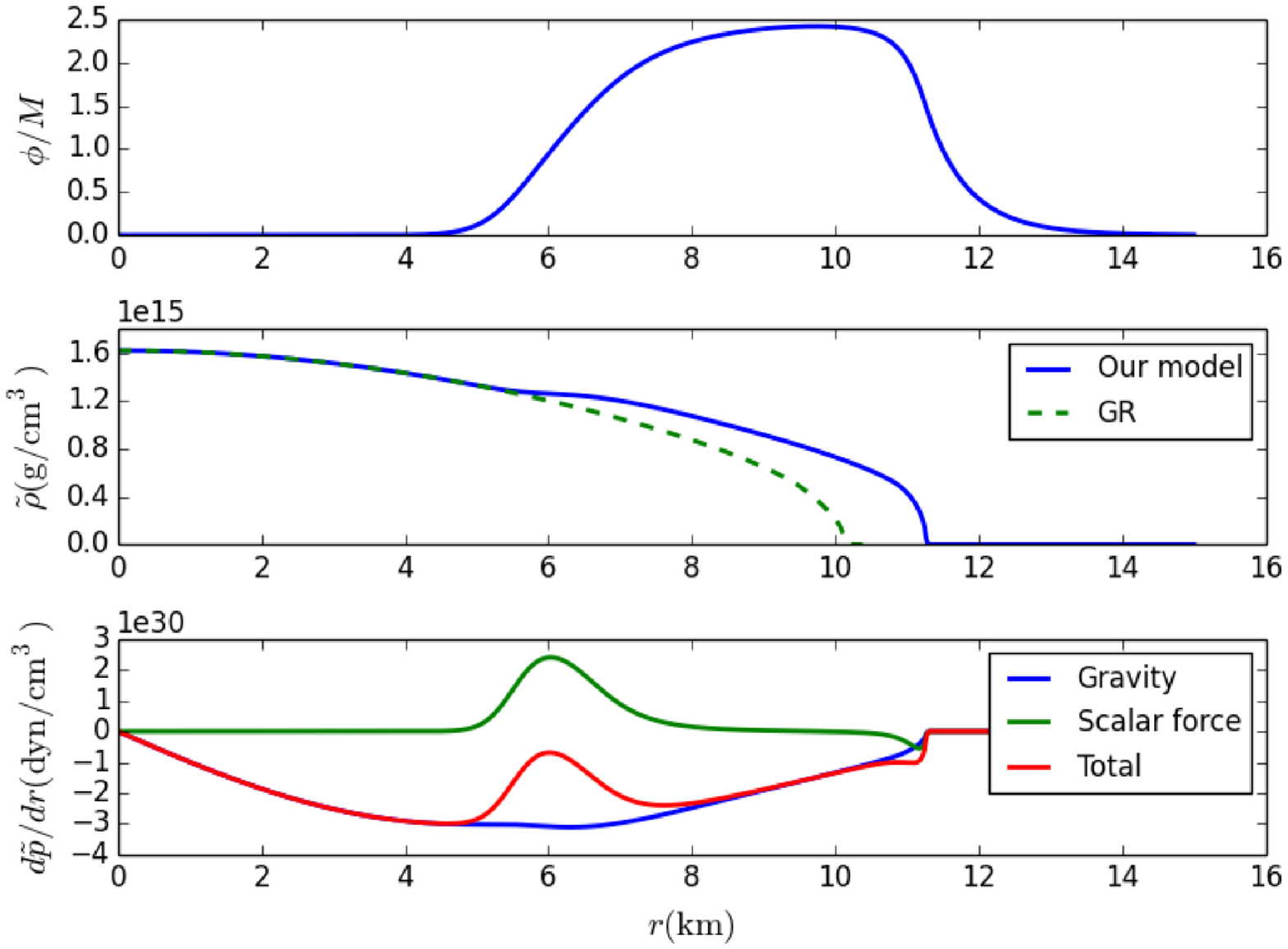}
             \end{center}
      \end{minipage} 

    \end{tabular}
    \caption{The profiles of $\phi$, $\tilde{\rho}$ and $d \tilde{p}/ d r$ for different central densities are shown. The upper one: $\tilde{\rho}(r=0) = 3 \times 10^{9}~\mathrm{MeV^4}=7.0 \times 10^{14}~\mathrm{g/cm^3}$, the lower one: $\tilde{\rho}(r=0)=7\times 10^{9}~\mathrm{MeV^4}=1.6 \times 10^{15}~\mathrm{g/cm^3}$. They have AP4 for the equation of state and $\eta=0.2,~\rhoPT=10^{8}~\mathrm{MeV^4}=2.3\times10^{13}~\mathrm{g/cm^3},~\lambda_\phi=1~\mathrm{km}$ for the parameters in our model. In the plot of $\tilde{\rho}(r)$, we also show the result of GR for the same central density (green dashed line). In the plot of $d \tilde{p}/ d r$, we plot the contribution of the gravity, $-(\tilde{\epsilon}+\tilde{p})\nu'/2$ (blue line), the scalar force,  $-(\tilde{\epsilon}+\tilde{p})d \mathrm{ln}A/dr$ (green line), and the total, $d \tilde{p}/dr$ (red line).}
    \label{profiles}
  \end{center}
\end{figure}

Next, we show the profiles of $\phi$, $\tilde{\rho}$ and $d \tilde{p}/ d r$ for different central densities in Fig. \ref{profiles}. For the lower central density, $\tilde{\rho}$ decreases more gradually than in GR because $\phi=\mathcal{O}(M)$ and the gravitational force is weakened. Near the surface, $\tilde{\rho}$ decreases drastically due to the compression by the scalar force. For the higher central density, the deviation of GR is extremely small in the core because $\phi \simeq 0$. In the region where $\phi$ grows as $r$ increases the decrease of $\tilde{\rho}$ becomes more gradual because the scalar force pushes the matter toward the outside. This effect also alters the internal structure significantly and such an effect of the scalar force is incorporated in the second term of the denominator of the right side in Eq. (\ref{peq2}) in the semianalytical approach. The gravitational force is weakened in the outer part of the star and the scalar force compresses the star near the surface, which is the same as that for the lower central density. 

In conclusion, there are three effects to alter the internal structure of the star in GR: the compression by the scalar force, the scalar force toward the outside for sufficiently high central densities, and the decrease of the effective gravitational constant. 
The first one decreases the mass of the star while the others increase it.

\subsection{The dependence of the mass-radius relation on $\eta,~\rhoPT$, and $\lambda_\phi$.}
\begin{figure}
  \begin{center}
    \begin{tabular}{c}

      \begin{minipage}{0.5\hsize}
        \begin{center}
          \includegraphics[clip, width=8cm]{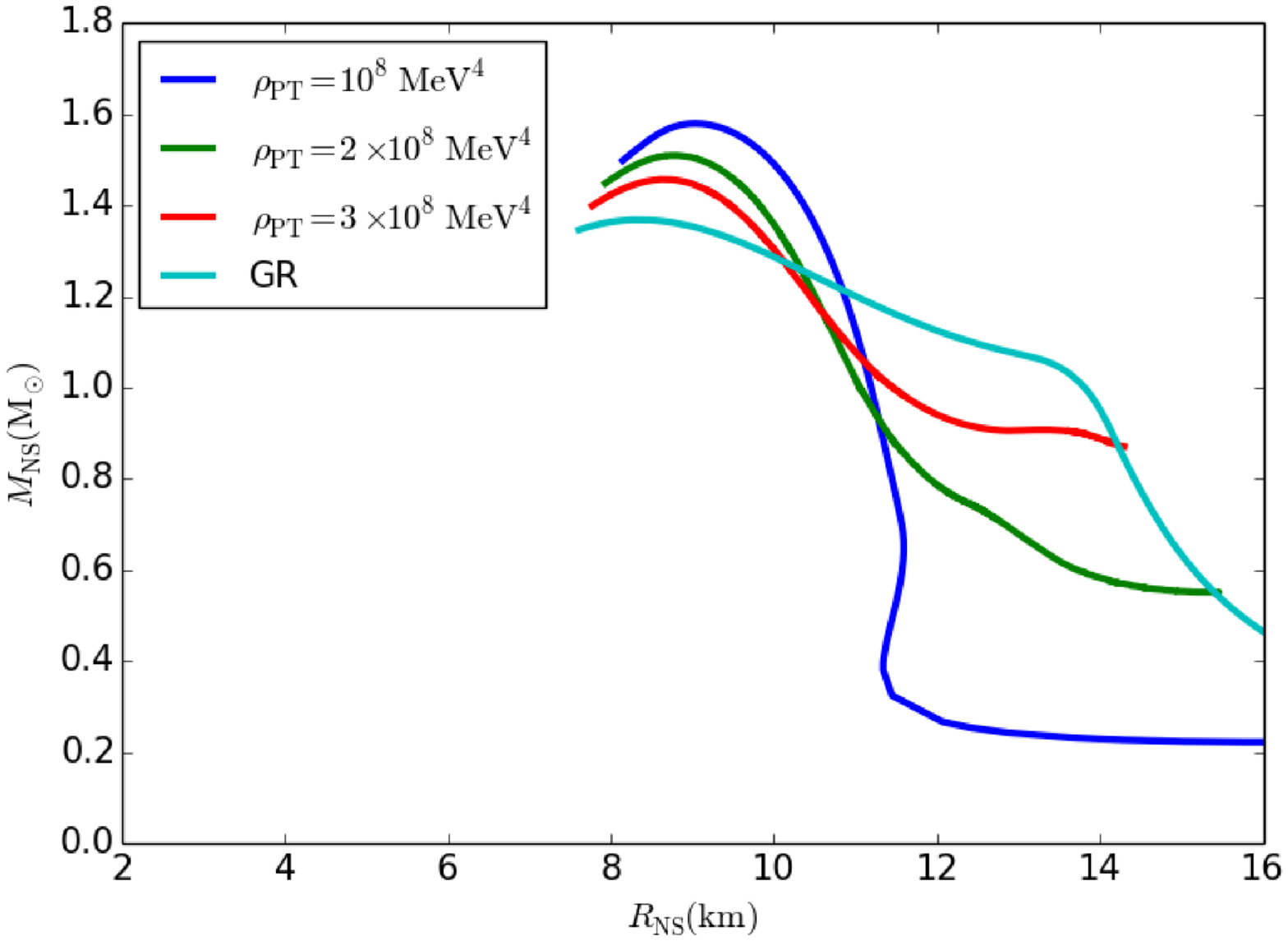}
        \end{center}
      \end{minipage} 

      \begin{minipage}{0.5\hsize}
 	      \begin{center}
    	     \includegraphics[clip, width=8cm]{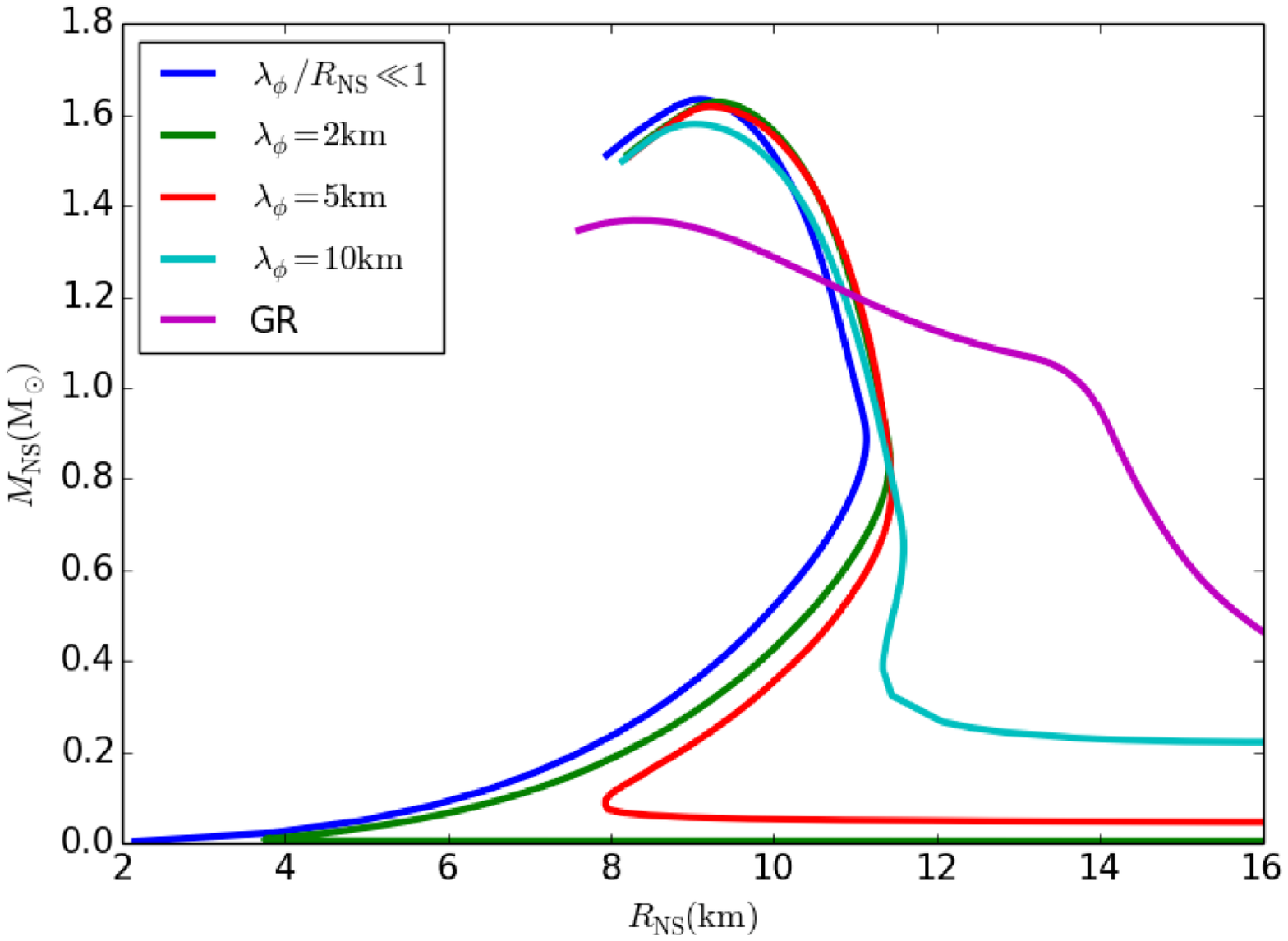}
        	\end{center}
      	\end{minipage}\\
      
            \begin{minipage}{0.5\hsize}
        \begin{center}
          \includegraphics[clip, width=8cm]{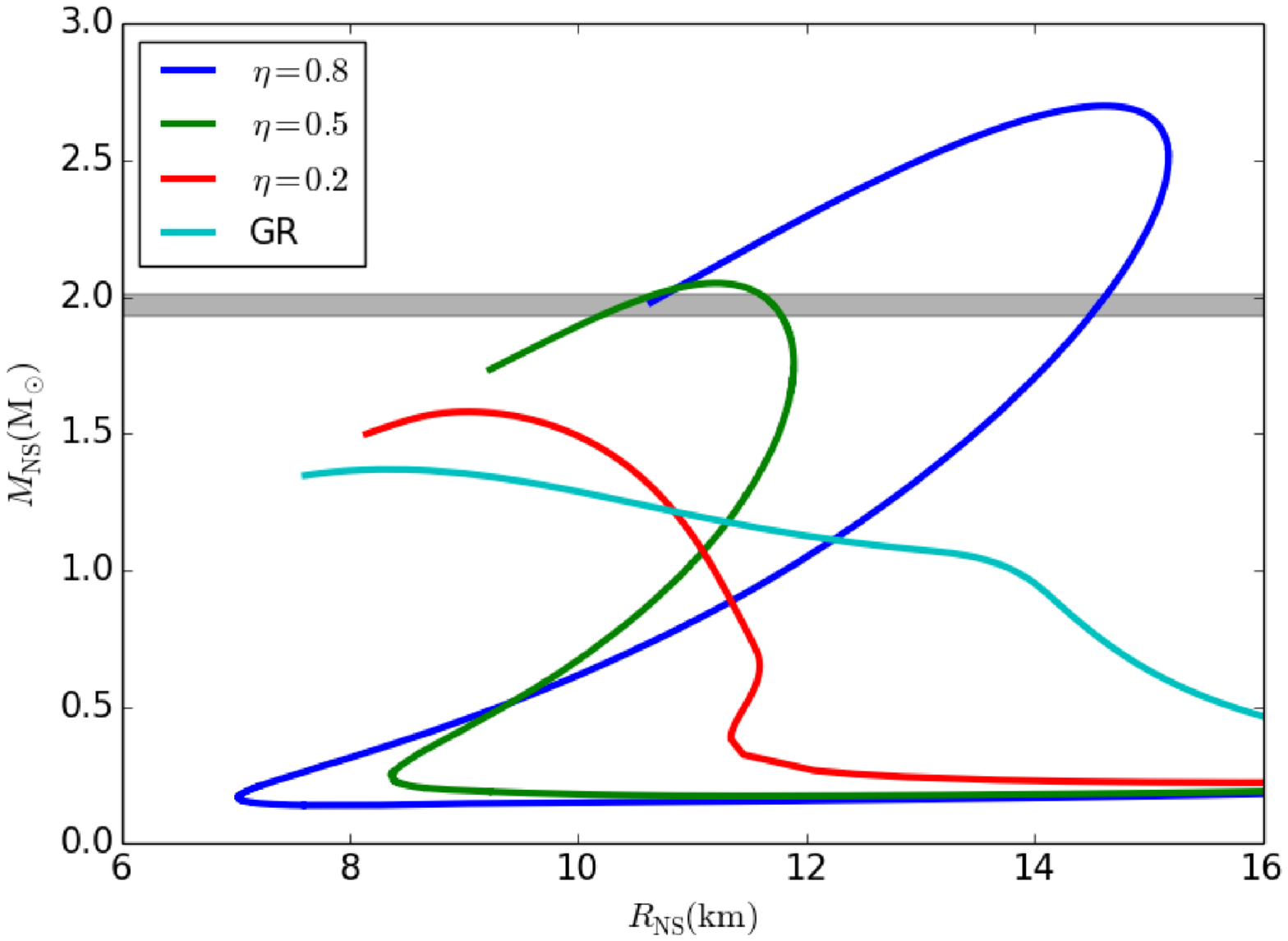}
        \end{center}
      \end{minipage}
            
    \end{tabular}
    \caption{The mass-radius relation curves for the GS1 are shown. $\RNS$ is the radius in the Einstein frame. The upper left one shows the dependence on $\rhoPT$ for $\eta=0.2,~\lambda_\phi=10~\mathrm{km}.$ The upper right one shows the dependence on $\lambda_\phi$ for $\eta=0.2,~\rhoPT=10^{8}~\mathrm{MeV^4}.$ The lower one shows the dependence on $\eta$ for $\rhoPT=10^{8}~\mathrm{MeV^4},~\lambda_\phi =10~\mathrm{km}.$ The shaded band shows the observational constraint from PSR J1614-2230 mass measurement of $1.97\pm0.04~M_{\odot}$.} 
    \label{MRGS1}
  \end{center}
\end{figure}

The dependence of the mass-radius relation on the parameters of our model for the GS1 is shown in Fig. \ref{MRGS1}.
Note that $\RNS$ is the radius in the Einstein frame and the physical surface area of the star is $4 \pi A^2(\phi(\RNS)) \RNS^2$.
In all cases, the maximum mass is larger than that in GR for a broad parameter region.
The result means that the two effects to increase the mass we explained are superior to the compression by the scalar force when the mass of the neutron star is around the maximum mass. 
The maximum mass increases as $\eta$ increases or $\rhoPT$ decreases. 
Especially, we find that the maximum mass reaches the mass of the massive pulsar PSR J1614-2230 even with the GS1 (the lower plot of Fig. \ref{MRGS1}).
Therefore, our model solves the inconsistency between the appearance of strange hadrons and the existence of the 2$M_\odot$ neutron stars.
On the other hand, the mass and radius are smaller than those in GR for lower central densities and the mass-radius curve approaches the origin.
This characteristic behavior is due to the compression by the scalar force.
This effect becomes more significant as $\lambda_\phi$ decreases.

\begin{figure}
  \begin{center}
    \begin{tabular}{c}

      \begin{minipage}{0.5\hsize}
        \begin{center}
          \includegraphics[clip, width=8cm]{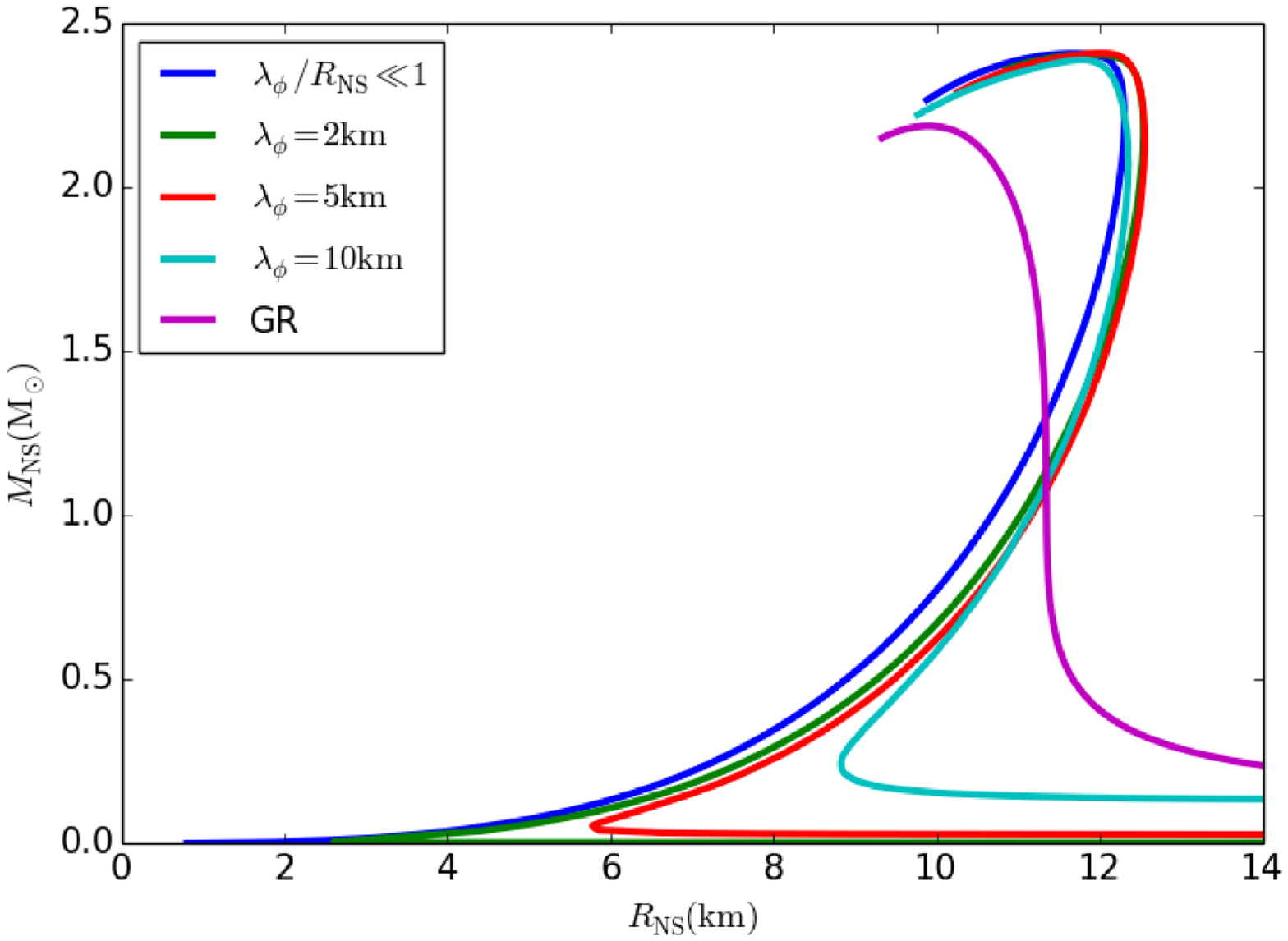}
           \end{center}
      \end{minipage}

      \begin{minipage}{0.5\hsize}
        \begin{center}
          \includegraphics[clip, width=8cm]{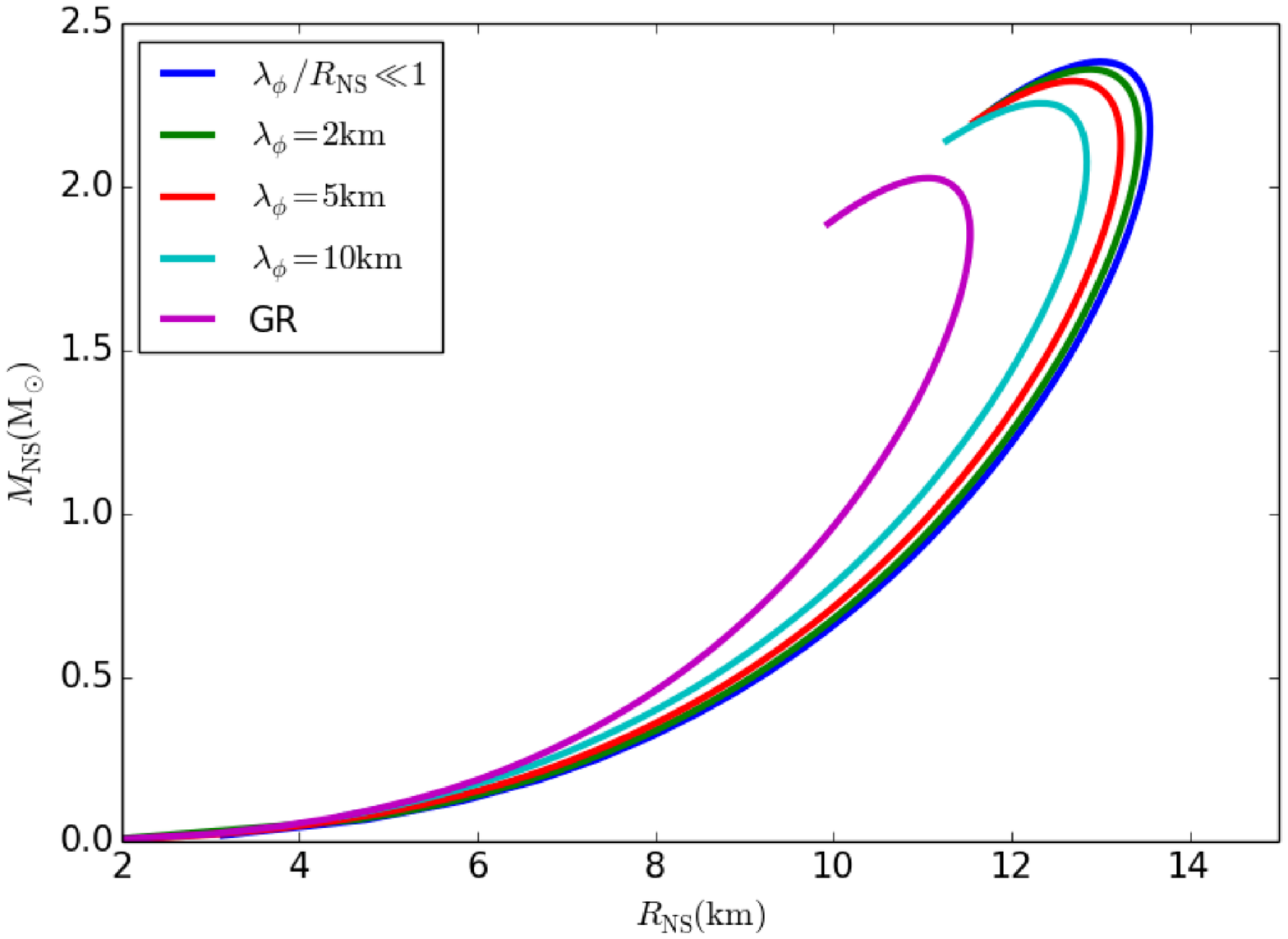}
            \end{center}
      \end{minipage} \\

    \end{tabular}
    \caption{The mass-radius relation curves for the AP4 (left) and MIT bag model (right) are shown for various values of $\lambda_\phi$. They have $\eta=0.2,~\rhoPT=10^8~\mathrm{MeV^4}=2.3 \times 10^{13}~\mathrm{g/cm^3}$ for the parameters in our model. $\RNS$ is the radius in the Einstein frame.}
    \label{EOSs}
  \end{center}
\end{figure}

The mass-radius relations for the other equations of state are shown in Fig. \ref{EOSs}. 
Its dependence on the parameters is qualitatively the same.

\subsection{The maximum mass}
\begin{figure}
  \begin{center}
    \begin{tabular}{c}

      \begin{minipage}{0.5\hsize}
        \begin{center}
          \includegraphics[clip, width=8cm]{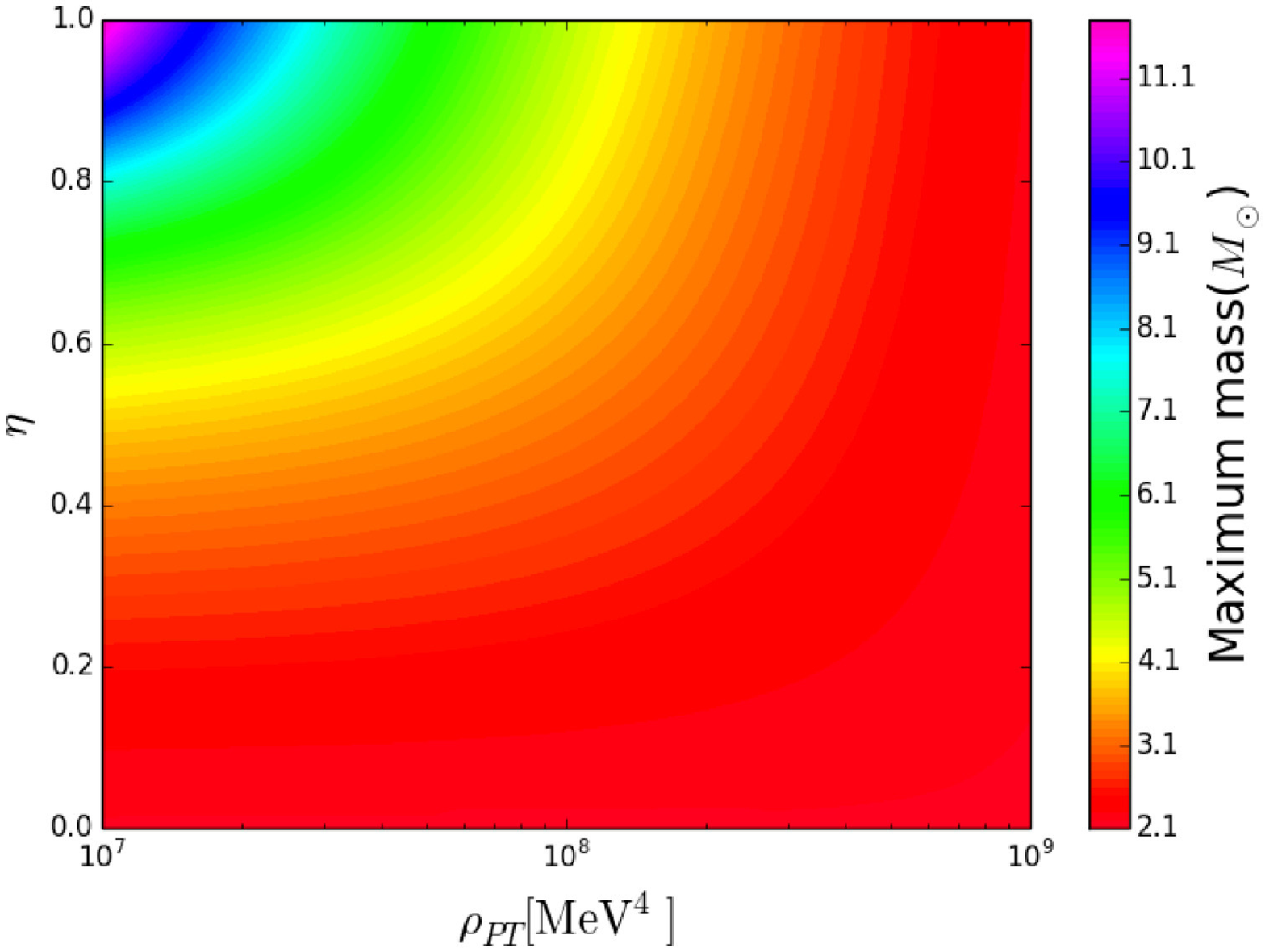}
           \end{center}
      \end{minipage}

      \begin{minipage}{0.5\hsize}
 	      \begin{center}
    	     \includegraphics[clip, width=8cm]{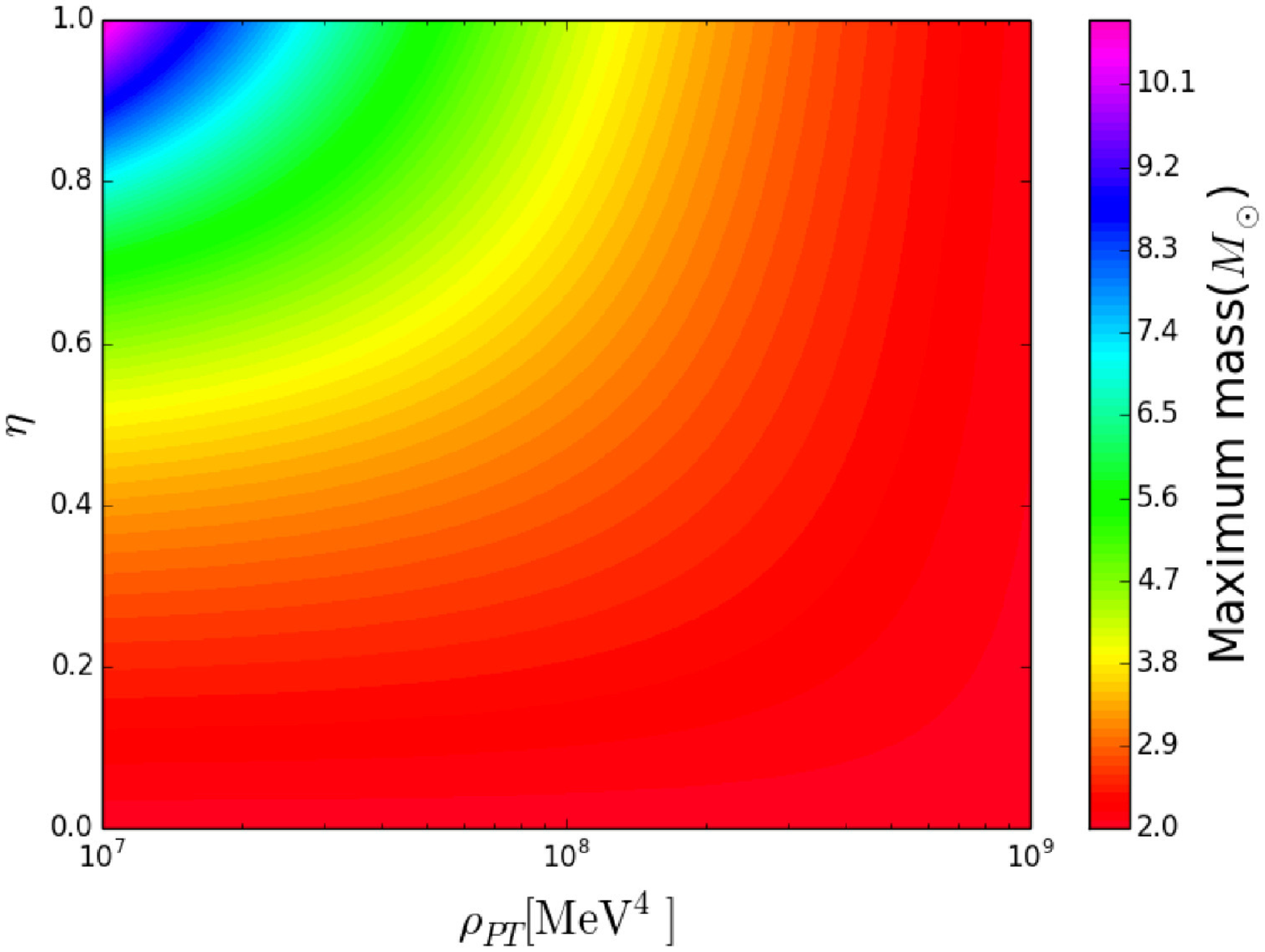}
	        	\end{center}
      	\end{minipage}\\
      
      \begin{minipage}{0.5\hsize}
        \begin{center}
          \includegraphics[clip, width=8cm]{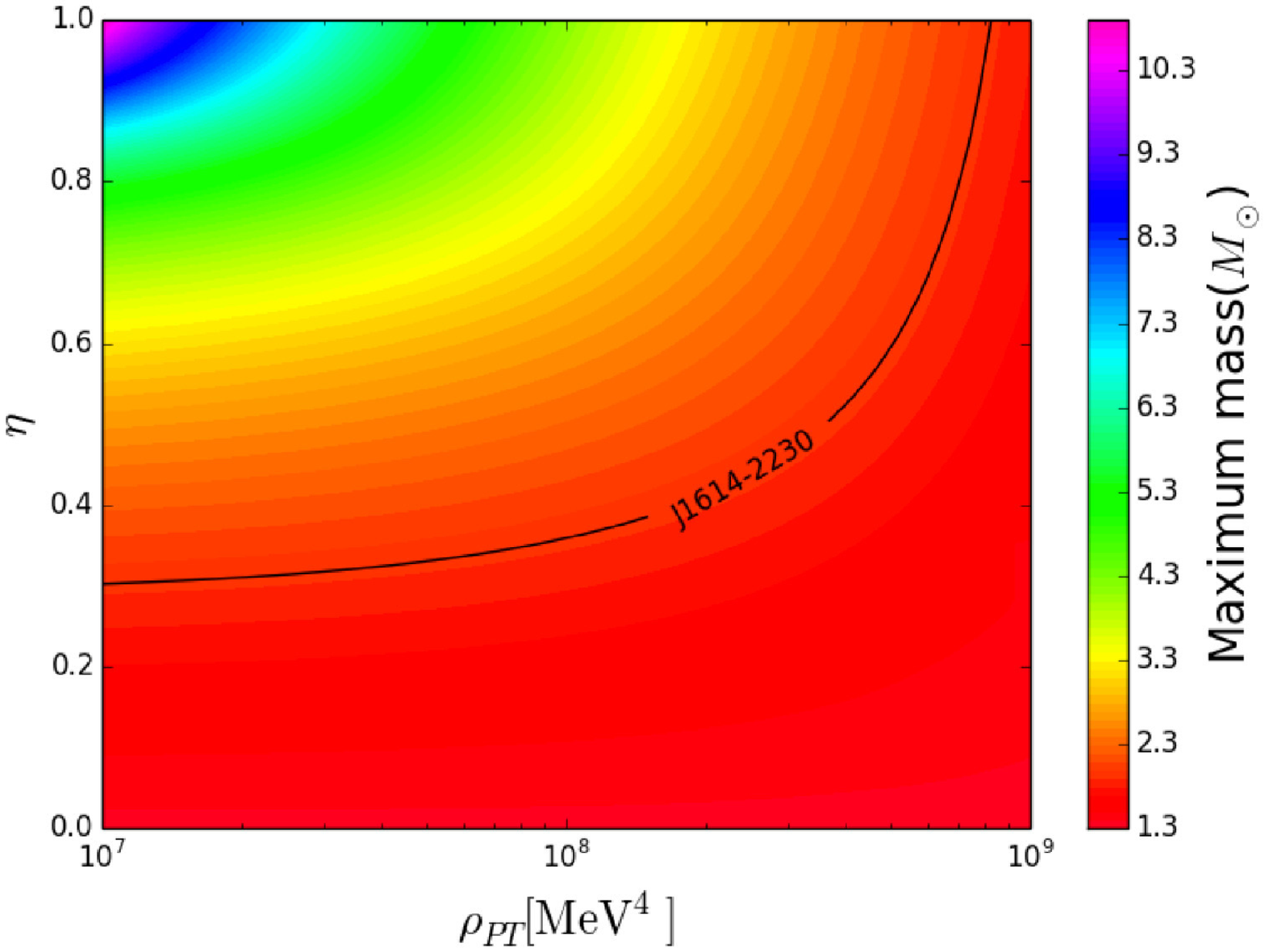}         
        \end{center}
      \end{minipage}
      
    \end{tabular}
    \caption{The maximum masses for AP4 (upper left), MIT bag model (upper right) and GS1 (lower) in the limit, $\lambda_\phi/\RNS \to 0$, are shown as functions of $\eta$ and $\rhoPT$. The black line in the lower plot shows the lower bound of the observational constraint from PSR J1614-2230 mass measurement, $1.93M_{\odot}$.}
    \label{contour}
  \end{center}
\end{figure}

We investigate the dependence of the maximum mass on the parameters of our model. 
We consider the case where $\lambda_\phi/\RNS \to 0$ and use the semianalytical approach to obtain the solution. 
In this case, the maximum mass is a function of $\eta$ and $\rhoPT$. 
The contour plot of the maximum mass is shown in Fig. \ref{contour}. 
There is a subtlety in obtaining this plot.
For some values of the parameters, the denominator of the right side in Eq. (\ref{peq2}) becomes zero in the region $r<\ri$.
In this case our semianalytical method is not applicable.
For example, such phenomena is observed in the low-$\eta$ region, $0.001 < \eta<0.02$, and the high-$\rhoPT$ region, $0.27<\eta<0.36,~6\times10^8~\mathrm{MeV^4}<\rhoPT<10^9~\mathrm{MeV^4}$, for the GS1. 
As can be seen in the lower plot of Fig. \ref{contour}, the maximum mass does not change so much from that in GR, which is not the case we are interested in.
The situation is the same for the other equations of state.
Therefore, we do not take this problem seriously and obtain the maximum mass for that parameter region through extrapolation.

Although the maximum mass for the GS1 in GR does not reach the lower bound of  the observational constraint from PSR J1614-2230 mass measurement, $1.93M_{\odot}$, it does for the values of the parameters in the left to the black line in Fig. \ref{contour} in our model.
In addition, we find that the maximum mass can also largely exceed the threshold that violates the causality bound in general relativity, which is $\sim 3M_{\odot}$ \cite{Rhoades:1974fn}.
The observational consequence of such extreme massive neutron stars is an interesting topic, but it is beyond the scope of this paper.

\section{Conclusions}\label{sec:conc}
We investigated the internal structure of neutron stars in a scalar-tensor theory, in which spontaneous scalarization by the massive scalar field occurs inside neutron stars.
In this paper, we focused on the case where the Compton wavelength of the scalar field $\lambda_\phi$ is shorter than $10~\mathrm{km}$.
In our model, spontaneous scalarization occurs and the effective gravitational constant decreases inside neutron stars while the scalar field decreases exponentially with the scale of $\lambda_\phi$ and GR is approximately correct outside the stars. 
In the mildly massive case, $10~\mathrm{km} \geq \lambda_\phi \gtrsim 1~\mathrm{km}$, we applied the shooting method to obtain solutions.
In the very massive case, $\lambda_\phi \ll 1~\mathrm{km}$, solving the equations numerically becomes extremely difficult and we developed a semianalytical method to obtain the solutions.

As a result of our analysis, we identified three important effects that alter the internal structure of neutron stars from that in GR: the compression by the scalar force, the scalar force toward the outside for sufficiently high central densities, and the decrease of the effective gravitational constant.
Especially, the effects of the scalar force are significant only when $\lambda_\phi \ll \RNS$.
We also found that the latter two are superior to the first one and the maximum mass is larger than in GR for a broad range of the parameters in our model.
Especially, the maximum mass can reach the mass of the massive pulsar PSR J1614-2230 even if strange hadrons are taken into account in our model.
In addition, it can largely exceed the threshold that violates the causality bound in general relativity.

There are some issues that, although important, have not been addressed in this paper. First is the stability of neutron stars in this theory. Since the scalar field sits on the stable point of the effective potential, we intuitively expect that the scalar field does not cause additional instability. It is worth checking this expectation. Second is the waveform of gravitational waves emitted by a coalescence of a binary, one of which is a neutron star in this theory. As we have shown in this paper, the scalar field significantly changes the internal structure of neutron stars. Such modification should propagate to the gravitational-wave waveform, for example, via the tidal deformability. Clarifying how the waveform is modified would be useful in testing the spontaneous-scalarization scenario. These are left for the future investigations.


\acknowledgements{
We thank Kipp Cannon, Yousuke Itoh, Kazumi Kashiyama, Shogo B. Kobayashi, Takashi Nakamura, Kazuhiro Nakazawa, Toshikazu Shigeyama, Kent Yagi and Jun'ichi Yokoyama for helpful discussions.
This work was supported by a research program of the Advanced Leading Graduate Course for Photon Science (ALPS) at the University of Tokyo (S.M.),
JSPS Grant-in-Aid for Young Scientists (B)
No.15K17632 (T.S.) and MEXT Grant-in-Aid for Scientific Research on
Innovative Areas No.17H06359 (T.S.).

}


\appendix
\section{The derivation of the WKB solution Equation (\ref{WKB1})}\label{sec:appendixA}
Because of Eq. (\ref{WKB_condition}), the nonlinear terms in Eq. (\ref{psieq}) are negligible and $\mu$ is approximately constant.
Therefore, we can approximate Eq. (\ref{psieq}) by
\begin{equation}
\frac{d^2 \phi}{d r^2} = -\frac{2}{r} \frac{1 - \frac{\mu}{r}}{1 - \frac{ 2 \mu}{r}} \frac{d \phi}{d r} + \frac{m^2_\phi}{1 - \frac{2 \mu}{r}}\phi. \label{phieq_B}
\end{equation}
We assume a WKB form for $\phi$,
\begin{equation}
\phi(r)=A(r) \mathrm{exp}[B(r)],
\end{equation}
with
\begin{equation}
\left| \frac{\frac{d A}{d r}}{A} \right| \sim \frac{1}{r}, ~ \left| \frac{\frac{d^2 B}{d r^2}}{ \left(\frac{d B}{d r} \right)^2} \right| \sim \frac{\lambda_\phi}{r} \ll 1. 
\end{equation}
Substituting this form into Eq. (\ref{phieq_B}) leads to
\begin{align}
\frac{d}{d r} \mathrm{ln} A &= \frac{3\mu - 2r}{r (r-2\mu)}, \\
\left(\frac{d B}{d r} \right)^2 &= \frac{m^2_\phi}{1 - \frac{2 \mu}{r}}.
\end{align}
Solving them, we obtain two independent solutions in Eq. (\ref{WKB1}). 

\section{The justification of Eq. (\ref{phi=phibar}) and Eq. (\ref{psi<<mphi})}\label{sec:appendixB}
In this section, we discuss the compatibility of Eq. (\ref{phi=phibar}) and Eq. (\ref{psi<<mphi}) in the region $r<\ri$ with the definition of $\ri$, Eq. (\ref{rs_def}). Since the deviation of $\phi$ from $\bar{\phi}$ grows exponentially, we can model $\phi(r)$ as
\begin{equation}
\phi(r)=\bar{\phi}(\tilde{p}(r))-\delta \phi_{\mathrm{i}} \mathrm{exp}\left[C\frac{r-\ri}{\lambda_\phi}\right],
\end{equation}
where $\delta \phi_{\mathrm{i}}$ is a constant and $C$ is an $\mathcal{O}(1)$ coefficient. 
Substituting this into Eq. (\ref{rs_def}) and taking into account only the terms first order in $\delta \phi_\mathrm{i}$, we obtain
\begin{equation}
\delta \phi_{\mathrm{i}} = -\frac{\lambda_\phi}{2 C \alpha(\bar{\phi}(\tilde{p}_{\mathrm{i}})) }\left(\frac{d \nu}{d r} -2 \frac{d \mathrm{ln} A(\bar{\phi}(\tilde{p}))}{d r}\right)_{r=r_{\mathrm{i}}}.
\end{equation}
Since $\alpha \simeq -1/M$, we can estimate $\delta \phi_{\mathrm{i}}$ as follows.
\begin{equation}
\delta \phi_{\mathrm{i}} \sim M\lambda_\phi \left(\frac{d \nu}{d r} -2 \frac{d \mathrm{ln} A(\bar{\phi}(\tilde{p}))}{d r}\right)_{r=r_{\mathrm{i}}} \lesssim M\frac{\lambda_\phi}{\RNS} \ll M.
\end{equation}
Therefore, the deviation from $\bar{\phi}$ is negligibly small and Eq. (\ref{phi=phibar}) can be applied in $0<r<r_{\mathrm{i}}$. In addition, $\delta \phi_\mathrm{i} /\lambda_\phi \ll m_\phi \bar{\phi}$. Therefore, Eq. (\ref{psi<<mphi}) can be also applied.

\section{The derivations of Eqs. (\ref{psieq3}) - (\ref{phi_integrate})}\label{sec:appendixC}
With Eq. (\ref{r<<l}) and Eq. (\ref{constants}), Eq. (\ref{psieq}) leads to
\begin{equation}
\begin{aligned}
(\ri - 2 \mui ) \frac{d^2 \phi}{d r^2} &= - 2 \left(1 - \frac{\mui}{\ri}\right)\frac{d \phi}{d r}+ m^2_\phi \left(4 \pi G \ri^2 \phi^2 \frac{d \phi}{d r} + \ri \phi \right)\\
&~~~~~~~~~~~~+\ri A^4(\phi)\left\{4 \pi G \ri (\tilde{\epsilon} - \tilde{p})\frac{d \phi}{d r} + \alpha (\tilde{\epsilon}-3\tilde{p}) \right\} \label{phieq4}
\end{aligned}
\end{equation}
Using $d \phi/d r \sim \phi/\lambda_\phi \sim M/\lambda_\phi$ and $d^2 \phi/dr^2 \sim \phi/\lambda^2_\phi \sim M/\lambda^2_\phi$, we have
\begin{align}
\left|\frac{\left(1-\frac{\mui}{\ri}\right)\frac{d \phi}{dr}}{m^2_\phi \ri \phi}\right| &\sim \frac{\lambda_\phi}{\ri} \ll 1,\\
\left| \frac{4 \pi G \ri^2 m^2_\phi \phi^2 \frac{d \phi}{dr}}{m^2_\phi \ri \phi} \right| &\sim 10^{-2} \eta \left(\frac{\ri}{10\mathrm{km}}\right)^2 \left(\frac{\rhoPT}{10^8\mathrm{MeV^4}}\right) \frac{\lambda_\phi}{\ri} \ll 1,\\
\left| \frac{ 4 \pi G \ri^2 A^4(\phi) (\tilde{\epsilon}-\tilde{p})\frac{d \phi}{d r}}{\ri \alpha A^4(\phi) (\tilde{\epsilon}-3 \tilde{p})} \right|&\sim 10^{-2}  \left(\frac{\ri}{10\mathrm{km}}\right)^2 \left(\frac{\rhoPT}{10^8\mathrm{MeV^4}}\right) \frac{\lambda_\phi}{\ri} \ll 1.
\end{align}
In the last line, we use the following estimation:
\begin{equation}
\frac{\eta}{2 M^2} \left| \frac{\tilde{\epsilon}-\tilde{p}}{\tilde{\epsilon}-3 \tilde{p}} \frac{\phi}{\alpha} \right| =
\left| \frac{\tilde{\epsilon}-\tilde{p}}{\tilde{\epsilon}-3 \tilde{p}} \right| \left(\eta  + (1 - \eta) \mathrm{exp}\left(\frac{\phi^2}{2 M^2}\right)\right) = \mathcal{O}(1),
\end{equation}
which holds true for the equations of state we use.
Therefore, the following three terms 
\begin{equation}
\left(1-\frac{\mui}{\ri}\right)\frac{d \phi}{dr},~4 \pi G \ri^2 m^2_\phi \phi^2 \frac{d \phi}{dr},~ 4 \pi G \ri^2 A^4(\phi) (\tilde{\epsilon}-\tilde{p})\frac{d \phi}{d r}
\end{equation}
are negligible compared to the other terms.
On the other hand, the ratio between the other remaining two terms is
\begin{equation}
\left| \frac{\ri \alpha A^4(\phi) (\tilde{\epsilon}-3 \tilde{p})}{m^2_\phi \ri \phi} \right| \sim \frac{2}{\eta} \frac{\tilde{\epsilon}-3 \tilde{p}}{\rhoPT}.
\end{equation} 
Therefore, they are comparable to each other and Eq. (\ref{phieq4}) can be reduced to Eq. (\ref{psieq3}).

Next, we explain how to derive Eq. (\ref{p_integrate}) and Eq. (\ref{phi_integrate}). 
Integrating Eq. (\ref{peq3}), we obtain
\begin{equation}
\int^{\pii}_{\tilde{p}} \frac{d \tilde{p}}{\tilde{\epsilon}+\tilde{p}} = \mathrm{ln}\frac{A}{A(\bar{\phi}(\pii))}. \label{integration_form}
\end{equation}
Using the first law of thermodynamics, Eq. (\ref{1stlaw}), we obtain
\begin{equation}
\int^{\pii}_{\tilde{p}} \frac{d \tilde{p}}{\tilde{\epsilon}+\tilde{p}} = \left[ \mathrm{ln}\left(\frac{\tilde{\epsilon}+\tilde{p}}{\tilde{\rho}}\right) \right]^{\pii}_{\tilde{p}}. 
\end{equation}
Substituting this into Eq. (\ref{integration_form}) leads to Eq. (\ref{p_integrate}). 
On the other hand, multiplying Eq. (\ref{psieq3}) by $d\phi/dr$ and integrating it from $r=\ri$, we obtain
\begin{equation}
\frac{1}{2} \psii^2 = \frac{1}{1-\frac{2 \mui}{\ri}} \left[\frac{1}{2} m^2_\phi (\phi^2 -\bar{\phi}^2(\pii))+ \int^{A}_{A(\bar{\phi}(\pii))} dA A^3 (\tilde{\epsilon}-3 \tilde{p}) \right], \label{C_3}
\end{equation}
where we ignore $\psii$ since $\psii \ll m_\phi \bar{\phi}(\pii)$.
Differentiating Eq. (\ref{integration_form}) with respect to $A$ leads to 
\begin{equation}
\frac{d \tilde{p}}{d A} = -\frac{\tilde{\epsilon}+\tilde{p}}{A}, 
\end{equation}
and we obtain
\begin{equation}
\frac{d}{d A}(\tilde{p} A^4) = -(\tilde{\epsilon}-3 \tilde{p}) A^3.
\end{equation}
Therefore, we can perform the integration in the right side of Eq. (\ref{C_3}) and obtain Eq. (\ref{phi_integrate}).


\bibliography{NSpaper}

\begin{thebibliography}{48}%
\makeatletter
\providecommand \@ifxundefined [1]{%
 \@ifx{#1\undefined}
}%
\providecommand \@ifnum [1]{%
 \ifnum #1\expandafter \@firstoftwo
 \else \expandafter \@secondoftwo
 \fi
}%
\providecommand \@ifx [1]{%
 \ifx #1\expandafter \@firstoftwo
 \else \expandafter \@secondoftwo
 \fi
}%
\providecommand \natexlab [1]{#1}%
\providecommand \enquote  [1]{``#1''}%
\providecommand \bibnamefont  [1]{#1}%
\providecommand \bibfnamefont [1]{#1}%
\providecommand \citenamefont [1]{#1}%
\providecommand \href@noop [0]{\@secondoftwo}%
\providecommand \href [0]{\begingroup \@sanitize@url \@href}%
\providecommand \@href[1]{\@@startlink{#1}\@@href}%
\providecommand \@@href[1]{\endgroup#1\@@endlink}%
\providecommand \@sanitize@url [0]{\catcode `\\12\catcode `\$12\catcode
  `\&12\catcode `\#12\catcode `\^12\catcode `\_12\catcode `\%12\relax}%
\providecommand \@@startlink[1]{}%
\providecommand \@@endlink[0]{}%
\providecommand \url  [0]{\begingroup\@sanitize@url \@url }%
\providecommand \@url [1]{\endgroup\@href {#1}{\urlprefix }}%
\providecommand \urlprefix  [0]{URL }%
\providecommand \Eprint [0]{\href }%
\providecommand \doibase [0]{http://dx.doi.org/}%
\providecommand \selectlanguage [0]{\@gobble}%
\providecommand \bibinfo  [0]{\@secondoftwo}%
\providecommand \bibfield  [0]{\@secondoftwo}%
\providecommand \translation [1]{[#1]}%
\providecommand \BibitemOpen [0]{}%
\providecommand \bibitemStop [0]{}%
\providecommand \bibitemNoStop [0]{.\EOS\space}%
\providecommand \EOS [0]{\spacefactor3000\relax}%
\providecommand \BibitemShut  [1]{\csname bibitem#1\endcsname}%
\let\auto@bib@innerbib\@empty
\bibitem [{\citenamefont {Will}(2014)}]{Will:2014kxa}%
  \BibitemOpen
  \bibfield  {author} {\bibinfo {author} {\bibfnamefont {C.~M.}\ \bibnamefont
  {Will}},\ }\href {\doibase 10.12942/lrr-2014-4} {\bibfield  {journal}
  {\bibinfo  {journal} {Living Rev. Rel.}\ }\textbf {\bibinfo {volume} {17}},\
  \bibinfo {pages} {4} (\bibinfo {year} {2014})},\ \Eprint
  {http://arxiv.org/abs/1403.7377} {arXiv:1403.7377 [gr-qc]} \BibitemShut
  {NoStop}%
\bibitem [{\citenamefont {Joyce}\ \emph {et~al.}(2015)\citenamefont {Joyce},
  \citenamefont {Jain}, \citenamefont {Khoury},\ and\ \citenamefont
  {Trodden}}]{Joyce:2014kja}%
  \BibitemOpen
  \bibfield  {author} {\bibinfo {author} {\bibfnamefont {A.}~\bibnamefont
  {Joyce}}, \bibinfo {author} {\bibfnamefont {B.}~\bibnamefont {Jain}},
  \bibinfo {author} {\bibfnamefont {J.}~\bibnamefont {Khoury}}, \ and\ \bibinfo
  {author} {\bibfnamefont {M.}~\bibnamefont {Trodden}},\ }\href {\doibase
  10.1016/j.physrep.2014.12.002} {\bibfield  {journal} {\bibinfo  {journal}
  {Phys. Rept.}\ }\textbf {\bibinfo {volume} {568}},\ \bibinfo {pages} {1}
  (\bibinfo {year} {2015})},\ \Eprint {http://arxiv.org/abs/1407.0059}
  {arXiv:1407.0059 [astro-ph.CO]} \BibitemShut {NoStop}%
\bibitem [{\citenamefont {Abbott}\ \emph
  {et~al.}(2016{\natexlab{a}})\citenamefont {Abbott} \emph
  {et~al.}}]{Abbott:2016blz}%
  \BibitemOpen
  \bibfield  {author} {\bibinfo {author} {\bibfnamefont {B.~P.}\ \bibnamefont
  {Abbott}} \emph {et~al.} (\bibinfo {collaboration} {Virgo, LIGO
  Scientific}),\ }\href {\doibase 10.1103/PhysRevLett.116.061102} {\bibfield
  {journal} {\bibinfo  {journal} {Phys. Rev. Lett.}\ }\textbf {\bibinfo
  {volume} {116}},\ \bibinfo {pages} {061102} (\bibinfo {year}
  {2016}{\natexlab{a}})},\ \Eprint {http://arxiv.org/abs/1602.03837}
  {arXiv:1602.03837 [gr-qc]} \BibitemShut {NoStop}%
\bibitem [{\citenamefont {Abbott}\ \emph
  {et~al.}(2016{\natexlab{b}})\citenamefont {Abbott} \emph
  {et~al.}}]{TheLIGOScientific:2016pea}%
  \BibitemOpen
  \bibfield  {author} {\bibinfo {author} {\bibfnamefont {B.~P.}\ \bibnamefont
  {Abbott}} \emph {et~al.} (\bibinfo {collaboration} {Virgo, LIGO
  Scientific}),\ }\href {\doibase 10.1103/PhysRevX.6.041015} {\bibfield
  {journal} {\bibinfo  {journal} {Phys. Rev.}\ }\textbf {\bibinfo {volume}
  {X6}},\ \bibinfo {pages} {041015} (\bibinfo {year} {2016}{\natexlab{b}})},\
  \Eprint {http://arxiv.org/abs/1606.04856} {arXiv:1606.04856 [gr-qc]}
  \BibitemShut {NoStop}%
\bibitem [{\citenamefont {Abbott}\ \emph {et~al.}(2017)\citenamefont {Abbott}
  \emph {et~al.}}]{Abbott:2017vtc}%
  \BibitemOpen
  \bibfield  {author} {\bibinfo {author} {\bibfnamefont {B.~P.}\ \bibnamefont
  {Abbott}} \emph {et~al.} (\bibinfo {collaboration} {VIRGO, LIGO
  Scientific}),\ }\href {\doibase 10.1103/PhysRevLett.118.221101} {\bibfield
  {journal} {\bibinfo  {journal} {Phys. Rev. Lett.}\ }\textbf {\bibinfo
  {volume} {118}},\ \bibinfo {pages} {221101} (\bibinfo {year} {2017})},\
  \Eprint {http://arxiv.org/abs/1706.01812} {arXiv:1706.01812 [gr-qc]}
  \BibitemShut {NoStop}%
\bibitem [{\citenamefont {Abbott}\ \emph
  {et~al.}(2016{\natexlab{c}})\citenamefont {Abbott} \emph
  {et~al.}}]{TheLIGOScientific:2016src}%
  \BibitemOpen
  \bibfield  {author} {\bibinfo {author} {\bibfnamefont {B.~P.}\ \bibnamefont
  {Abbott}} \emph {et~al.} (\bibinfo {collaboration} {Virgo, LIGO
  Scientific}),\ }\href {\doibase 10.1103/PhysRevLett.116.221101} {\bibfield
  {journal} {\bibinfo  {journal} {Phys. Rev. Lett.}\ }\textbf {\bibinfo
  {volume} {116}},\ \bibinfo {pages} {221101} (\bibinfo {year}
  {2016}{\natexlab{c}})},\ \Eprint {http://arxiv.org/abs/1602.03841}
  {arXiv:1602.03841 [gr-qc]} \BibitemShut {NoStop}%
\bibitem [{\citenamefont {Yunes}\ \emph {et~al.}(2016)\citenamefont {Yunes},
  \citenamefont {Yagi},\ and\ \citenamefont {Pretorius}}]{Yunes:2016jcc}%
  \BibitemOpen
  \bibfield  {author} {\bibinfo {author} {\bibfnamefont {N.}~\bibnamefont
  {Yunes}}, \bibinfo {author} {\bibfnamefont {K.}~\bibnamefont {Yagi}}, \ and\
  \bibinfo {author} {\bibfnamefont {F.}~\bibnamefont {Pretorius}},\ }\href
  {\doibase 10.1103/PhysRevD.94.084002} {\bibfield  {journal} {\bibinfo
  {journal} {Phys. Rev.}\ }\textbf {\bibinfo {volume} {D94}},\ \bibinfo {pages}
  {084002} (\bibinfo {year} {2016})},\ \Eprint
  {http://arxiv.org/abs/1603.08955} {arXiv:1603.08955 [gr-qc]} \BibitemShut
  {NoStop}%
\bibitem [{\citenamefont {Berti}\ \emph {et~al.}(2015)\citenamefont {Berti}
  \emph {et~al.}}]{Berti:2015itd}%
  \BibitemOpen
  \bibfield  {author} {\bibinfo {author} {\bibfnamefont {E.}~\bibnamefont
  {Berti}} \emph {et~al.},\ }\href {\doibase 10.1088/0264-9381/32/24/243001}
  {\bibfield  {journal} {\bibinfo  {journal} {Class. Quant. Grav.}\ }\textbf
  {\bibinfo {volume} {32}},\ \bibinfo {pages} {243001} (\bibinfo {year}
  {2015})},\ \Eprint {http://arxiv.org/abs/1501.07274} {arXiv:1501.07274
  [gr-qc]} \BibitemShut {NoStop}%
\bibitem [{\citenamefont {Babichev}\ \emph {et~al.}(2016)\citenamefont
  {Babichev}, \citenamefont {Koyama}, \citenamefont {Langlois}, \citenamefont
  {Saito},\ and\ \citenamefont {Sakstein}}]{Babichev:2016jom}%
  \BibitemOpen
  \bibfield  {author} {\bibinfo {author} {\bibfnamefont {E.}~\bibnamefont
  {Babichev}}, \bibinfo {author} {\bibfnamefont {K.}~\bibnamefont {Koyama}},
  \bibinfo {author} {\bibfnamefont {D.}~\bibnamefont {Langlois}}, \bibinfo
  {author} {\bibfnamefont {R.}~\bibnamefont {Saito}}, \ and\ \bibinfo {author}
  {\bibfnamefont {J.}~\bibnamefont {Sakstein}},\ }\href {\doibase
  10.1088/0264-9381/33/23/235014} {\bibfield  {journal} {\bibinfo  {journal}
  {Class. Quant. Grav.}\ }\textbf {\bibinfo {volume} {33}},\ \bibinfo {pages}
  {235014} (\bibinfo {year} {2016})},\ \Eprint
  {http://arxiv.org/abs/1606.06627} {arXiv:1606.06627 [gr-qc]} \BibitemShut
  {NoStop}%
\bibitem [{\citenamefont {Sakstein}\ \emph {et~al.}(2017)\citenamefont
  {Sakstein}, \citenamefont {Babichev}, \citenamefont {Koyama}, \citenamefont
  {Langlois},\ and\ \citenamefont {Saito}}]{Sakstein:2016oel}%
  \BibitemOpen
  \bibfield  {author} {\bibinfo {author} {\bibfnamefont {J.}~\bibnamefont
  {Sakstein}}, \bibinfo {author} {\bibfnamefont {E.}~\bibnamefont {Babichev}},
  \bibinfo {author} {\bibfnamefont {K.}~\bibnamefont {Koyama}}, \bibinfo
  {author} {\bibfnamefont {D.}~\bibnamefont {Langlois}}, \ and\ \bibinfo
  {author} {\bibfnamefont {R.}~\bibnamefont {Saito}},\ }\href {\doibase
  10.1103/PhysRevD.95.064013} {\bibfield  {journal} {\bibinfo  {journal} {Phys.
  Rev.}\ }\textbf {\bibinfo {volume} {D95}},\ \bibinfo {pages} {064013}
  (\bibinfo {year} {2017})},\ \Eprint {http://arxiv.org/abs/1612.04263}
  {arXiv:1612.04263 [gr-qc]} \BibitemShut {NoStop}%
\bibitem [{\citenamefont {Minamitsuji}\ and\ \citenamefont
  {Silva}(2016)}]{Minamitsuji:2016hkk}%
  \BibitemOpen
  \bibfield  {author} {\bibinfo {author} {\bibfnamefont {M.}~\bibnamefont
  {Minamitsuji}}\ and\ \bibinfo {author} {\bibfnamefont {H.~O.}\ \bibnamefont
  {Silva}},\ }\href {\doibase 10.1103/PhysRevD.93.124041} {\bibfield  {journal}
  {\bibinfo  {journal} {Phys. Rev.}\ }\textbf {\bibinfo {volume} {D93}},\
  \bibinfo {pages} {124041} (\bibinfo {year} {2016})},\ \Eprint
  {http://arxiv.org/abs/1604.07742} {arXiv:1604.07742 [gr-qc]} \BibitemShut
  {NoStop}%
\bibitem [{\citenamefont {Maselli}\ \emph {et~al.}(2016)\citenamefont
  {Maselli}, \citenamefont {Silva}, \citenamefont {Minamitsuji},\ and\
  \citenamefont {Berti}}]{Maselli:2016gxk}%
  \BibitemOpen
  \bibfield  {author} {\bibinfo {author} {\bibfnamefont {A.}~\bibnamefont
  {Maselli}}, \bibinfo {author} {\bibfnamefont {H.~O.}\ \bibnamefont {Silva}},
  \bibinfo {author} {\bibfnamefont {M.}~\bibnamefont {Minamitsuji}}, \ and\
  \bibinfo {author} {\bibfnamefont {E.}~\bibnamefont {Berti}},\ }\href
  {\doibase 10.1103/PhysRevD.93.124056} {\bibfield  {journal} {\bibinfo
  {journal} {Phys. Rev.}\ }\textbf {\bibinfo {volume} {D93}},\ \bibinfo {pages}
  {124056} (\bibinfo {year} {2016})},\ \Eprint
  {http://arxiv.org/abs/1603.04876} {arXiv:1603.04876 [gr-qc]} \BibitemShut
  {NoStop}%
\bibitem [{\citenamefont {Aoki}\ \emph {et~al.}(2016)\citenamefont {Aoki},
  \citenamefont {Maeda},\ and\ \citenamefont {Tanabe}}]{Aoki:2016eov}%
  \BibitemOpen
  \bibfield  {author} {\bibinfo {author} {\bibfnamefont {K.}~\bibnamefont
  {Aoki}}, \bibinfo {author} {\bibfnamefont {K.-i.}\ \bibnamefont {Maeda}}, \
  and\ \bibinfo {author} {\bibfnamefont {M.}~\bibnamefont {Tanabe}},\ }\href
  {\doibase 10.1103/PhysRevD.93.064054} {\bibfield  {journal} {\bibinfo
  {journal} {Phys. Rev.}\ }\textbf {\bibinfo {volume} {D93}},\ \bibinfo {pages}
  {064054} (\bibinfo {year} {2016})},\ \Eprint
  {http://arxiv.org/abs/1602.02227} {arXiv:1602.02227 [gr-qc]} \BibitemShut
  {NoStop}%
\bibitem [{\citenamefont {Cisterna}\ \emph {et~al.}(2015)\citenamefont
  {Cisterna}, \citenamefont {Delsate},\ and\ \citenamefont
  {Rinaldi}}]{Cisterna:2015yla}%
  \BibitemOpen
  \bibfield  {author} {\bibinfo {author} {\bibfnamefont {A.}~\bibnamefont
  {Cisterna}}, \bibinfo {author} {\bibfnamefont {T.}~\bibnamefont {Delsate}}, \
  and\ \bibinfo {author} {\bibfnamefont {M.}~\bibnamefont {Rinaldi}},\ }\href
  {\doibase 10.1103/PhysRevD.92.044050} {\bibfield  {journal} {\bibinfo
  {journal} {Phys. Rev.}\ }\textbf {\bibinfo {volume} {D92}},\ \bibinfo {pages}
  {044050} (\bibinfo {year} {2015})},\ \Eprint
  {http://arxiv.org/abs/1504.05189} {arXiv:1504.05189 [gr-qc]} \BibitemShut
  {NoStop}%
\bibitem [{\citenamefont {Cisterna}\ \emph {et~al.}(2016)\citenamefont
  {Cisterna}, \citenamefont {Delsate}, \citenamefont {Ducobu},\ and\
  \citenamefont {Rinaldi}}]{Cisterna:2016vdx}%
  \BibitemOpen
  \bibfield  {author} {\bibinfo {author} {\bibfnamefont {A.}~\bibnamefont
  {Cisterna}}, \bibinfo {author} {\bibfnamefont {T.}~\bibnamefont {Delsate}},
  \bibinfo {author} {\bibfnamefont {L.}~\bibnamefont {Ducobu}}, \ and\ \bibinfo
  {author} {\bibfnamefont {M.}~\bibnamefont {Rinaldi}},\ }\href {\doibase
  10.1103/PhysRevD.93.084046} {\bibfield  {journal} {\bibinfo  {journal} {Phys.
  Rev.}\ }\textbf {\bibinfo {volume} {D93}},\ \bibinfo {pages} {084046}
  (\bibinfo {year} {2016})},\ \Eprint {http://arxiv.org/abs/1602.06939}
  {arXiv:1602.06939 [gr-qc]} \BibitemShut {NoStop}%
\bibitem [{\citenamefont {Postnikov}\ \emph {et~al.}(2010)\citenamefont
  {Postnikov}, \citenamefont {Prakash},\ and\ \citenamefont
  {Lattimer}}]{Postnikov:2010yn}%
  \BibitemOpen
  \bibfield  {author} {\bibinfo {author} {\bibfnamefont {S.}~\bibnamefont
  {Postnikov}}, \bibinfo {author} {\bibfnamefont {M.}~\bibnamefont {Prakash}},
  \ and\ \bibinfo {author} {\bibfnamefont {J.~M.}\ \bibnamefont {Lattimer}},\
  }\href {\doibase 10.1103/PhysRevD.82.024016} {\bibfield  {journal} {\bibinfo
  {journal} {Phys. Rev.}\ }\textbf {\bibinfo {volume} {D82}},\ \bibinfo {pages}
  {024016} (\bibinfo {year} {2010})},\ \Eprint {http://arxiv.org/abs/1004.5098}
  {arXiv:1004.5098 [astro-ph.SR]} \BibitemShut {NoStop}%
\bibitem [{\citenamefont {Hinderer}\ \emph {et~al.}(2010)\citenamefont
  {Hinderer}, \citenamefont {Lackey}, \citenamefont {Lang},\ and\ \citenamefont
  {Read}}]{Hinderer:2009ca}%
  \BibitemOpen
  \bibfield  {author} {\bibinfo {author} {\bibfnamefont {T.}~\bibnamefont
  {Hinderer}}, \bibinfo {author} {\bibfnamefont {B.~D.}\ \bibnamefont
  {Lackey}}, \bibinfo {author} {\bibfnamefont {R.~N.}\ \bibnamefont {Lang}}, \
  and\ \bibinfo {author} {\bibfnamefont {J.~S.}\ \bibnamefont {Read}},\ }\href
  {\doibase 10.1103/PhysRevD.81.123016} {\bibfield  {journal} {\bibinfo
  {journal} {Phys. Rev.}\ }\textbf {\bibinfo {volume} {D81}},\ \bibinfo {pages}
  {123016} (\bibinfo {year} {2010})},\ \Eprint {http://arxiv.org/abs/0911.3535}
  {arXiv:0911.3535 [astro-ph.HE]} \BibitemShut {NoStop}%
\bibitem [{\citenamefont {Yagi}\ and\ \citenamefont
  {Yunes}(2013)}]{Yagi:2013bca}%
  \BibitemOpen
  \bibfield  {author} {\bibinfo {author} {\bibfnamefont {K.}~\bibnamefont
  {Yagi}}\ and\ \bibinfo {author} {\bibfnamefont {N.}~\bibnamefont {Yunes}},\
  }\href {\doibase 10.1126/science.1236462} {\bibfield  {journal} {\bibinfo
  {journal} {Science}\ }\textbf {\bibinfo {volume} {341}},\ \bibinfo {pages}
  {365} (\bibinfo {year} {2013})},\ \Eprint {http://arxiv.org/abs/1302.4499}
  {arXiv:1302.4499 [gr-qc]} \BibitemShut {NoStop}%
\bibitem [{\citenamefont {Flanagan}\ and\ \citenamefont
  {Hinderer}(2008)}]{Flanagan:2007ix}%
  \BibitemOpen
  \bibfield  {author} {\bibinfo {author} {\bibfnamefont {E.~E.}\ \bibnamefont
  {Flanagan}}\ and\ \bibinfo {author} {\bibfnamefont {T.}~\bibnamefont
  {Hinderer}},\ }\href {\doibase 10.1103/PhysRevD.77.021502} {\bibfield
  {journal} {\bibinfo  {journal} {Phys. Rev.}\ }\textbf {\bibinfo {volume}
  {D77}},\ \bibinfo {pages} {021502} (\bibinfo {year} {2008})},\ \Eprint
  {http://arxiv.org/abs/0709.1915} {arXiv:0709.1915 [astro-ph]} \BibitemShut
  {NoStop}%
\bibitem [{\citenamefont {Demorest}\ \emph {et~al.}(2010)\citenamefont
  {Demorest}, \citenamefont {Pennucci}, \citenamefont {Ransom}, \citenamefont
  {Roberts},\ and\ \citenamefont {Hessels}}]{Demorest:2010bx}%
  \BibitemOpen
  \bibfield  {author} {\bibinfo {author} {\bibfnamefont {P.}~\bibnamefont
  {Demorest}}, \bibinfo {author} {\bibfnamefont {T.}~\bibnamefont {Pennucci}},
  \bibinfo {author} {\bibfnamefont {S.}~\bibnamefont {Ransom}}, \bibinfo
  {author} {\bibfnamefont {M.}~\bibnamefont {Roberts}}, \ and\ \bibinfo
  {author} {\bibfnamefont {J.}~\bibnamefont {Hessels}},\ }\href {\doibase
  10.1038/nature09466} {\bibfield  {journal} {\bibinfo  {journal} {Nature}\
  }\textbf {\bibinfo {volume} {467}},\ \bibinfo {pages} {1081} (\bibinfo {year}
  {2010})},\ \Eprint {http://arxiv.org/abs/1010.5788} {arXiv:1010.5788
  [astro-ph.HE]} \BibitemShut {NoStop}%
\bibitem [{\citenamefont {Antoniadis}\ \emph {et~al.}(2013)\citenamefont
  {Antoniadis} \emph {et~al.}}]{Antoniadis:2013pzd}%
  \BibitemOpen
  \bibfield  {author} {\bibinfo {author} {\bibfnamefont {J.}~\bibnamefont
  {Antoniadis}} \emph {et~al.},\ }\href {\doibase 10.1126/science.1233232}
  {\bibfield  {journal} {\bibinfo  {journal} {Science}\ }\textbf {\bibinfo
  {volume} {340}},\ \bibinfo {pages} {6131} (\bibinfo {year} {2013})},\ \Eprint
  {http://arxiv.org/abs/1304.6875} {arXiv:1304.6875 [astro-ph.HE]} \BibitemShut
  {NoStop}%
\bibitem [{\citenamefont {Nishizaki}\ \emph {et~al.}(2002)\citenamefont
  {Nishizaki}, \citenamefont {Takatsuka},\ and\ \citenamefont
  {Yamamoto}}]{Nishizaki:2002ih}%
  \BibitemOpen
  \bibfield  {author} {\bibinfo {author} {\bibfnamefont {S.}~\bibnamefont
  {Nishizaki}}, \bibinfo {author} {\bibfnamefont {T.}~\bibnamefont
  {Takatsuka}}, \ and\ \bibinfo {author} {\bibfnamefont {Y.}~\bibnamefont
  {Yamamoto}},\ }\href {\doibase 10.1143/PTP.108.703} {\bibfield  {journal}
  {\bibinfo  {journal} {Prog. Theor. Phys.}\ }\textbf {\bibinfo {volume}
  {108}},\ \bibinfo {pages} {703} (\bibinfo {year} {2002})}\BibitemShut
  {NoStop}%
\bibitem [{\citenamefont {Gandolfi}\ \emph {et~al.}(2015)\citenamefont
  {Gandolfi}, \citenamefont {Gezerlis},\ and\ \citenamefont
  {Carlson}}]{Gandolfi:2015jma}%
  \BibitemOpen
  \bibfield  {author} {\bibinfo {author} {\bibfnamefont {S.}~\bibnamefont
  {Gandolfi}}, \bibinfo {author} {\bibfnamefont {A.}~\bibnamefont {Gezerlis}},
  \ and\ \bibinfo {author} {\bibfnamefont {J.}~\bibnamefont {Carlson}},\ }\href
  {\doibase 10.1146/annurev-nucl-102014-021957} {\bibfield  {journal} {\bibinfo
   {journal} {Ann. Rev. Nucl. Part. Sci.}\ }\textbf {\bibinfo {volume} {65}},\
  \bibinfo {pages} {303} (\bibinfo {year} {2015})},\ \Eprint
  {http://arxiv.org/abs/1501.05675} {arXiv:1501.05675 [nucl-th]} \BibitemShut
  {NoStop}%
\bibitem [{\citenamefont {Rikovska-Stone}\ \emph {et~al.}(2007)\citenamefont
  {Rikovska-Stone}, \citenamefont {Guichon}, \citenamefont {Matevosyan},\ and\
  \citenamefont {Thomas}}]{RikovskaStone:2006ta}%
  \BibitemOpen
  \bibfield  {author} {\bibinfo {author} {\bibfnamefont {J.}~\bibnamefont
  {Rikovska-Stone}}, \bibinfo {author} {\bibfnamefont {P.~A.~M.}\ \bibnamefont
  {Guichon}}, \bibinfo {author} {\bibfnamefont {H.~H.}\ \bibnamefont
  {Matevosyan}}, \ and\ \bibinfo {author} {\bibfnamefont {A.~W.}\ \bibnamefont
  {Thomas}},\ }\href {\doibase 10.1016/j.nuclphysa.2007.05.011} {\bibfield
  {journal} {\bibinfo  {journal} {Nucl. Phys.}\ }\textbf {\bibinfo {volume}
  {A792}},\ \bibinfo {pages} {341} (\bibinfo {year} {2007})},\ \Eprint
  {http://arxiv.org/abs/nucl-th/0611030} {arXiv:nucl-th/0611030 [nucl-th]}
  \BibitemShut {NoStop}%
\bibitem [{\citenamefont {Bonanno}\ and\ \citenamefont
  {Sedrakian}(2012)}]{Bonanno:2011ch}%
  \BibitemOpen
  \bibfield  {author} {\bibinfo {author} {\bibfnamefont {L.}~\bibnamefont
  {Bonanno}}\ and\ \bibinfo {author} {\bibfnamefont {A.}~\bibnamefont
  {Sedrakian}},\ }\href {\doibase 10.1051/0004-6361/201117832} {\bibfield
  {journal} {\bibinfo  {journal} {Astron. Astrophys.}\ }\textbf {\bibinfo
  {volume} {539}},\ \bibinfo {pages} {A16} (\bibinfo {year} {2012})},\ \Eprint
  {http://arxiv.org/abs/1108.0559} {arXiv:1108.0559 [astro-ph.SR]} \BibitemShut
  {NoStop}%
\bibitem [{\citenamefont {Masuda}\ \emph {et~al.}(2013)\citenamefont {Masuda},
  \citenamefont {Hatsuda},\ and\ \citenamefont {Takatsuka}}]{Masuda:2012ed}%
  \BibitemOpen
  \bibfield  {author} {\bibinfo {author} {\bibfnamefont {K.}~\bibnamefont
  {Masuda}}, \bibinfo {author} {\bibfnamefont {T.}~\bibnamefont {Hatsuda}}, \
  and\ \bibinfo {author} {\bibfnamefont {T.}~\bibnamefont {Takatsuka}},\ }\href
  {\doibase 10.1093/ptep/ptt045} {\bibfield  {journal} {\bibinfo  {journal}
  {PTEP}\ }\textbf {\bibinfo {volume} {2013}},\ \bibinfo {pages} {073D01}
  (\bibinfo {year} {2013})},\ \Eprint {http://arxiv.org/abs/1212.6803}
  {arXiv:1212.6803 [nucl-th]} \BibitemShut {NoStop}%
\bibitem [{\citenamefont {Astashenok}\ \emph {et~al.}(2014)\citenamefont
  {Astashenok}, \citenamefont {Capozziello},\ and\ \citenamefont
  {Odintsov}}]{Astashenok:2014pua}%
  \BibitemOpen
  \bibfield  {author} {\bibinfo {author} {\bibfnamefont {A.~V.}\ \bibnamefont
  {Astashenok}}, \bibinfo {author} {\bibfnamefont {S.}~\bibnamefont
  {Capozziello}}, \ and\ \bibinfo {author} {\bibfnamefont {S.~D.}\ \bibnamefont
  {Odintsov}},\ }\href {\doibase 10.1103/PhysRevD.89.103509} {\bibfield
  {journal} {\bibinfo  {journal} {Phys. Rev.}\ }\textbf {\bibinfo {volume}
  {D89}},\ \bibinfo {pages} {103509} (\bibinfo {year} {2014})},\ \Eprint
  {http://arxiv.org/abs/1401.4546} {arXiv:1401.4546 [gr-qc]} \BibitemShut
  {NoStop}%
\bibitem [{\citenamefont {Fujii}\ and\ \citenamefont
  {Maeda}(2007)}]{Fujii:2003pa}%
  \BibitemOpen
  \bibfield  {author} {\bibinfo {author} {\bibfnamefont {Y.}~\bibnamefont
  {Fujii}}\ and\ \bibinfo {author} {\bibfnamefont {K.}~\bibnamefont {Maeda}},\
  }\href {http://www.cambridge.org/uk/catalogue/catalogue.asp?isbn=0521811597}
  {\emph {\bibinfo {title} {{The scalar-tensor theory of gravitation}}}}\
  (\bibinfo  {publisher} {Cambridge University Press},\ \bibinfo {year}
  {2007})\BibitemShut {NoStop}%
\bibitem [{\citenamefont {Bertotti}\ \emph {et~al.}(2003)\citenamefont
  {Bertotti}, \citenamefont {Iess},\ and\ \citenamefont
  {Tortora}}]{Bertotti:2003rm}%
  \BibitemOpen
  \bibfield  {author} {\bibinfo {author} {\bibfnamefont {B.}~\bibnamefont
  {Bertotti}}, \bibinfo {author} {\bibfnamefont {L.}~\bibnamefont {Iess}}, \
  and\ \bibinfo {author} {\bibfnamefont {P.}~\bibnamefont {Tortora}},\ }\href
  {\doibase 10.1038/nature01997} {\bibfield  {journal} {\bibinfo  {journal}
  {Nature}\ }\textbf {\bibinfo {volume} {425}},\ \bibinfo {pages} {374}
  (\bibinfo {year} {2003})}\BibitemShut {NoStop}%
\bibitem [{\citenamefont {Damour}\ and\ \citenamefont
  {Esposito-Farese}(1993)}]{Damour:1993hw}%
  \BibitemOpen
  \bibfield  {author} {\bibinfo {author} {\bibfnamefont {T.}~\bibnamefont
  {Damour}}\ and\ \bibinfo {author} {\bibfnamefont {G.}~\bibnamefont
  {Esposito-Farese}},\ }\href {\doibase 10.1103/PhysRevLett.70.2220} {\bibfield
   {journal} {\bibinfo  {journal} {Phys. Rev. Lett.}\ }\textbf {\bibinfo
  {volume} {70}},\ \bibinfo {pages} {2220} (\bibinfo {year}
  {1993})}\BibitemShut {NoStop}%
\bibitem [{\citenamefont {Damour}\ and\ \citenamefont
  {Esposito-Farese}(1996)}]{Damour:1996ke}%
  \BibitemOpen
  \bibfield  {author} {\bibinfo {author} {\bibfnamefont {T.}~\bibnamefont
  {Damour}}\ and\ \bibinfo {author} {\bibfnamefont {G.}~\bibnamefont
  {Esposito-Farese}},\ }\href {\doibase 10.1103/PhysRevD.54.1474} {\bibfield
  {journal} {\bibinfo  {journal} {Phys. Rev.}\ }\textbf {\bibinfo {volume}
  {D54}},\ \bibinfo {pages} {1474} (\bibinfo {year} {1996})},\ \Eprint
  {http://arxiv.org/abs/gr-qc/9602056} {arXiv:gr-qc/9602056 [gr-qc]}
  \BibitemShut {NoStop}%
\bibitem [{\citenamefont {Sampson}\ \emph {et~al.}(2014)\citenamefont
  {Sampson}, \citenamefont {Yunes}, \citenamefont {Cornish}, \citenamefont
  {Ponce}, \citenamefont {Barausse}, \citenamefont {Klein}, \citenamefont
  {Palenzuela},\ and\ \citenamefont {Lehner}}]{Sampson:2014qqa}%
  \BibitemOpen
  \bibfield  {author} {\bibinfo {author} {\bibfnamefont {L.}~\bibnamefont
  {Sampson}}, \bibinfo {author} {\bibfnamefont {N.}~\bibnamefont {Yunes}},
  \bibinfo {author} {\bibfnamefont {N.}~\bibnamefont {Cornish}}, \bibinfo
  {author} {\bibfnamefont {M.}~\bibnamefont {Ponce}}, \bibinfo {author}
  {\bibfnamefont {E.}~\bibnamefont {Barausse}}, \bibinfo {author}
  {\bibfnamefont {A.}~\bibnamefont {Klein}}, \bibinfo {author} {\bibfnamefont
  {C.}~\bibnamefont {Palenzuela}}, \ and\ \bibinfo {author} {\bibfnamefont
  {L.}~\bibnamefont {Lehner}},\ }\href {\doibase 10.1103/PhysRevD.90.124091}
  {\bibfield  {journal} {\bibinfo  {journal} {Phys. Rev.}\ }\textbf {\bibinfo
  {volume} {D90}},\ \bibinfo {pages} {124091} (\bibinfo {year} {2014})},\
  \Eprint {http://arxiv.org/abs/1407.7038} {arXiv:1407.7038 [gr-qc]}
  \BibitemShut {NoStop}%
\bibitem [{\citenamefont {Damour}\ and\ \citenamefont
  {Nordtvedt}(1993{\natexlab{a}})}]{Damour:1992kf}%
  \BibitemOpen
  \bibfield  {author} {\bibinfo {author} {\bibfnamefont {T.}~\bibnamefont
  {Damour}}\ and\ \bibinfo {author} {\bibfnamefont {K.}~\bibnamefont
  {Nordtvedt}},\ }\href {\doibase 10.1103/PhysRevLett.70.2217} {\bibfield
  {journal} {\bibinfo  {journal} {Phys. Rev. Lett.}\ }\textbf {\bibinfo
  {volume} {70}},\ \bibinfo {pages} {2217} (\bibinfo {year}
  {1993}{\natexlab{a}})}\BibitemShut {NoStop}%
\bibitem [{\citenamefont {Damour}\ and\ \citenamefont
  {Nordtvedt}(1993{\natexlab{b}})}]{Damour:1993id}%
  \BibitemOpen
  \bibfield  {author} {\bibinfo {author} {\bibfnamefont {T.}~\bibnamefont
  {Damour}}\ and\ \bibinfo {author} {\bibfnamefont {K.}~\bibnamefont
  {Nordtvedt}},\ }\href {\doibase 10.1103/PhysRevD.48.3436} {\bibfield
  {journal} {\bibinfo  {journal} {Phys. Rev.}\ }\textbf {\bibinfo {volume}
  {D48}},\ \bibinfo {pages} {3436} (\bibinfo {year}
  {1993}{\natexlab{b}})}\BibitemShut {NoStop}%
\bibitem [{\citenamefont {Chen}\ \emph {et~al.}(2015)\citenamefont {Chen},
  \citenamefont {Suyama},\ and\ \citenamefont {Yokoyama}}]{Chen:2015zmx}%
  \BibitemOpen
  \bibfield  {author} {\bibinfo {author} {\bibfnamefont {P.}~\bibnamefont
  {Chen}}, \bibinfo {author} {\bibfnamefont {T.}~\bibnamefont {Suyama}}, \ and\
  \bibinfo {author} {\bibfnamefont {J.}~\bibnamefont {Yokoyama}},\ }\href
  {\doibase 10.1103/PhysRevD.92.124016} {\bibfield  {journal} {\bibinfo
  {journal} {Phys. Rev.}\ }\textbf {\bibinfo {volume} {D92}},\ \bibinfo {pages}
  {124016} (\bibinfo {year} {2015})},\ \Eprint
  {http://arxiv.org/abs/1508.01384} {arXiv:1508.01384 [gr-qc]} \BibitemShut
  {NoStop}%
\bibitem [{\citenamefont {Ramazano\u{g}lu}\ and\ \citenamefont
  {Pretorius}(2016)}]{Ramazanoglu:2016kul}%
  \BibitemOpen
  \bibfield  {author} {\bibinfo {author} {\bibfnamefont {F.~M.}\ \bibnamefont
  {Ramazano\u{g}lu}}\ and\ \bibinfo {author} {\bibfnamefont {F.}~\bibnamefont
  {Pretorius}},\ }\href {\doibase 10.1103/PhysRevD.93.064005} {\bibfield
  {journal} {\bibinfo  {journal} {Phys. Rev.}\ }\textbf {\bibinfo {volume}
  {D93}},\ \bibinfo {pages} {064005} (\bibinfo {year} {2016})},\ \Eprint
  {http://arxiv.org/abs/1601.07475} {arXiv:1601.07475 [gr-qc]} \BibitemShut
  {NoStop}%
\bibitem [{\citenamefont {Khoury}\ and\ \citenamefont
  {Weltman}(2004)}]{Khoury:2003rn}%
  \BibitemOpen
  \bibfield  {author} {\bibinfo {author} {\bibfnamefont {J.}~\bibnamefont
  {Khoury}}\ and\ \bibinfo {author} {\bibfnamefont {A.}~\bibnamefont
  {Weltman}},\ }\href {\doibase 10.1103/PhysRevD.69.044026} {\bibfield
  {journal} {\bibinfo  {journal} {Phys. Rev.}\ }\textbf {\bibinfo {volume}
  {D69}},\ \bibinfo {pages} {044026} (\bibinfo {year} {2004})},\ \Eprint
  {http://arxiv.org/abs/astro-ph/0309411} {arXiv:astro-ph/0309411 [astro-ph]}
  \BibitemShut {NoStop}%
\bibitem [{\citenamefont {Itoh}(1970)}]{Itoh:1970uw}%
  \BibitemOpen
  \bibfield  {author} {\bibinfo {author} {\bibfnamefont {N.}~\bibnamefont
  {Itoh}},\ }\href {\doibase 10.1143/PTP.44.291} {\bibfield  {journal}
  {\bibinfo  {journal} {Prog. Theor. Phys.}\ }\textbf {\bibinfo {volume}
  {44}},\ \bibinfo {pages} {291} (\bibinfo {year} {1970})}\BibitemShut
  {NoStop}%
\bibitem [{\citenamefont {Witten}(1984)}]{Witten:1984rs}%
  \BibitemOpen
  \bibfield  {author} {\bibinfo {author} {\bibfnamefont {E.}~\bibnamefont
  {Witten}},\ }\href {\doibase 10.1103/PhysRevD.30.272} {\bibfield  {journal}
  {\bibinfo  {journal} {Phys. Rev.}\ }\textbf {\bibinfo {volume} {D30}},\
  \bibinfo {pages} {272} (\bibinfo {year} {1984})}\BibitemShut {NoStop}%
\bibitem [{\citenamefont {Haensel}\ \emph {et~al.}(1986)\citenamefont
  {Haensel}, \citenamefont {Zdunik},\ and\ \citenamefont
  {Schaeffer}}]{Haensel:1986qb}%
  \BibitemOpen
  \bibfield  {author} {\bibinfo {author} {\bibfnamefont {P.}~\bibnamefont
  {Haensel}}, \bibinfo {author} {\bibfnamefont {J.~L.}\ \bibnamefont {Zdunik}},
  \ and\ \bibinfo {author} {\bibfnamefont {R.}~\bibnamefont {Schaeffer}},\
  }\href@noop {} {\bibfield  {journal} {\bibinfo  {journal} {Astron.
  Astrophys.}\ }\textbf {\bibinfo {volume} {160}},\ \bibinfo {pages} {121}
  (\bibinfo {year} {1986})}\BibitemShut {NoStop}%
\bibitem [{\citenamefont {Alcock}\ \emph {et~al.}(1986)\citenamefont {Alcock},
  \citenamefont {Farhi},\ and\ \citenamefont {Olinto}}]{Alcock:1986hz}%
  \BibitemOpen
  \bibfield  {author} {\bibinfo {author} {\bibfnamefont {C.}~\bibnamefont
  {Alcock}}, \bibinfo {author} {\bibfnamefont {E.}~\bibnamefont {Farhi}}, \
  and\ \bibinfo {author} {\bibfnamefont {A.}~\bibnamefont {Olinto}},\ }\href
  {\doibase 10.1086/164679} {\bibfield  {journal} {\bibinfo  {journal}
  {Astrophys. J.}\ }\textbf {\bibinfo {volume} {310}},\ \bibinfo {pages} {261}
  (\bibinfo {year} {1986})}\BibitemShut {NoStop}%
\bibitem [{\citenamefont {Olinto}(1987)}]{Olinto:1986je}%
  \BibitemOpen
  \bibfield  {author} {\bibinfo {author} {\bibfnamefont {A.~V.}\ \bibnamefont
  {Olinto}},\ }\href {\doibase 10.1016/0370-2693(87)91144-0} {\bibfield
  {journal} {\bibinfo  {journal} {Phys. Lett.}\ }\textbf {\bibinfo {volume}
  {B192}},\ \bibinfo {pages} {71} (\bibinfo {year} {1987})}\BibitemShut
  {NoStop}%
\bibitem [{\citenamefont {Read}\ \emph {et~al.}(2009)\citenamefont {Read},
  \citenamefont {Lackey}, \citenamefont {Owen},\ and\ \citenamefont
  {Friedman}}]{Read:2008iy}%
  \BibitemOpen
  \bibfield  {author} {\bibinfo {author} {\bibfnamefont {J.~S.}\ \bibnamefont
  {Read}}, \bibinfo {author} {\bibfnamefont {B.~D.}\ \bibnamefont {Lackey}},
  \bibinfo {author} {\bibfnamefont {B.~J.}\ \bibnamefont {Owen}}, \ and\
  \bibinfo {author} {\bibfnamefont {J.~L.}\ \bibnamefont {Friedman}},\ }\href
  {\doibase 10.1103/PhysRevD.79.124032} {\bibfield  {journal} {\bibinfo
  {journal} {Phys. Rev.}\ }\textbf {\bibinfo {volume} {D79}},\ \bibinfo {pages}
  {124032} (\bibinfo {year} {2009})},\ \Eprint {http://arxiv.org/abs/0812.2163}
  {arXiv:0812.2163 [astro-ph]} \BibitemShut {NoStop}%
\bibitem [{\citenamefont {Akmal}\ \emph {et~al.}(1998)\citenamefont {Akmal},
  \citenamefont {Pandharipande},\ and\ \citenamefont
  {Ravenhall}}]{Akmal:1998cf}%
  \BibitemOpen
  \bibfield  {author} {\bibinfo {author} {\bibfnamefont {A.}~\bibnamefont
  {Akmal}}, \bibinfo {author} {\bibfnamefont {V.~R.}\ \bibnamefont
  {Pandharipande}}, \ and\ \bibinfo {author} {\bibfnamefont {D.~G.}\
  \bibnamefont {Ravenhall}},\ }\href {\doibase 10.1103/PhysRevC.58.1804}
  {\bibfield  {journal} {\bibinfo  {journal} {Phys. Rev.}\ }\textbf {\bibinfo
  {volume} {C58}},\ \bibinfo {pages} {1804} (\bibinfo {year} {1998})},\ \Eprint
  {http://arxiv.org/abs/nucl-th/9804027} {arXiv:nucl-th/9804027 [nucl-th]}
  \BibitemShut {NoStop}%
\bibitem [{\citenamefont {Glendenning}\ and\ \citenamefont
  {Schaffner-Bielich}(1999)}]{Glendenning:1997ak}%
  \BibitemOpen
  \bibfield  {author} {\bibinfo {author} {\bibfnamefont {N.~K.}\ \bibnamefont
  {Glendenning}}\ and\ \bibinfo {author} {\bibfnamefont {J.}~\bibnamefont
  {Schaffner-Bielich}},\ }\href {\doibase 10.1103/PhysRevC.60.025803}
  {\bibfield  {journal} {\bibinfo  {journal} {Phys. Rev.}\ }\textbf {\bibinfo
  {volume} {C60}},\ \bibinfo {pages} {025803} (\bibinfo {year} {1999})},\
  \Eprint {http://arxiv.org/abs/astro-ph/9810290} {arXiv:astro-ph/9810290
  [astro-ph]} \BibitemShut {NoStop}%
\bibitem [{\citenamefont {Chodos}\ \emph {et~al.}(1974)\citenamefont {Chodos},
  \citenamefont {Jaffe}, \citenamefont {Johnson}, \citenamefont {Thorn},\ and\
  \citenamefont {Weisskopf}}]{Chodos:1974je}%
  \BibitemOpen
  \bibfield  {author} {\bibinfo {author} {\bibfnamefont {A.}~\bibnamefont
  {Chodos}}, \bibinfo {author} {\bibfnamefont {R.~L.}\ \bibnamefont {Jaffe}},
  \bibinfo {author} {\bibfnamefont {K.}~\bibnamefont {Johnson}}, \bibinfo
  {author} {\bibfnamefont {C.~B.}\ \bibnamefont {Thorn}}, \ and\ \bibinfo
  {author} {\bibfnamefont {V.~F.}\ \bibnamefont {Weisskopf}},\ }\href {\doibase
  10.1103/PhysRevD.9.3471} {\bibfield  {journal} {\bibinfo  {journal} {Phys.
  Rev.}\ }\textbf {\bibinfo {volume} {D9}},\ \bibinfo {pages} {3471} (\bibinfo
  {year} {1974})}\BibitemShut {NoStop}%
\bibitem [{\citenamefont {Lattimer}(2012)}]{Lattimer:2012nd}%
  \BibitemOpen
  \bibfield  {author} {\bibinfo {author} {\bibfnamefont {J.~M.}\ \bibnamefont
  {Lattimer}},\ }\href {\doibase 10.1146/annurev-nucl-102711-095018} {\bibfield
   {journal} {\bibinfo  {journal} {Ann. Rev. Nucl. Part. Sci.}\ }\textbf
  {\bibinfo {volume} {62}},\ \bibinfo {pages} {485} (\bibinfo {year} {2012})},\
  \Eprint {http://arxiv.org/abs/1305.3510} {arXiv:1305.3510 [nucl-th]}
  \BibitemShut {NoStop}%
\bibitem [{\citenamefont {Rhoades}\ and\ \citenamefont
  {Ruffini}(1974)}]{Rhoades:1974fn}%
  \BibitemOpen
  \bibfield  {author} {\bibinfo {author} {\bibfnamefont {C.~E.}\ \bibnamefont
  {Rhoades}, \bibfnamefont {Jr.}}\ and\ \bibinfo {author} {\bibfnamefont
  {R.}~\bibnamefont {Ruffini}},\ }\href {\doibase 10.1103/PhysRevLett.32.324}
  {\bibfield  {journal} {\bibinfo  {journal} {Phys. Rev. Lett.}\ }\textbf
  {\bibinfo {volume} {32}},\ \bibinfo {pages} {324} (\bibinfo {year}
  {1974})}\BibitemShut {NoStop}%
\end{thebibliography}%

\end{document}